\documentclass[prd,showpacs,superscriptaddress,nofootinbib,floatfix,11pt]{revtex4-2}

\usepackage[utf8]{inputenc}
\usepackage[T1]{fontenc}
\usepackage{textgreek}
\usepackage{amsmath,amssymb,amsfonts,bm}
\usepackage{slashed}
\usepackage{subfigure}
\usepackage[colorlinks=true,
breaklinks=true,
urlcolor=magenta,
citecolor=blue]{hyperref}
\usepackage{appendix}
\usepackage{braket,bm}
\usepackage{multirow}
\usepackage{enumitem}
\usepackage[usenames,dvipsnames]{xcolor}
\usepackage{slashed}
\usepackage{orcidlink}
\usepackage{graphicx}
\usepackage{tikz}
\usepackage{tikz-feynman}
\usepackage{footnotebackref}
\usepackage{booktabs}
\usepackage{array}
\usepackage{overpic}
\usepackage{float}
\usepackage{threeparttable}
\allowdisplaybreaks[4]
\usetikzlibrary{arrows.meta}

\newcommand{\be}{\begin{equation}} 
\newcommand{\ee}{\end{equation}}
\newcommand{\bea}{\begin{eqnarray}} 
\newcommand{\eea}{\end{eqnarray}}
\newcommand{\beas}{\begin{eqnarray*}} 
\newcommand{\eeas}{\end{eqnarray*}}

\newcommand{\dast}{$D_0^*(2300)$}

\newcommand{\uestc}{\affiliation{School of Physics, University of Electronic Science and Technology of China, Chengdu 611731, China}}
\newcommand{\ific}{\affiliation{Instituto de F\'isica Corpuscular (centro mixto CSIC-UV), \\
Institutos de Investigaci\'on de Paterna, Apartado 22085, 46071, Valencia, Spain}}

\begin{document}

\title{\boldmath Analysis of the $D_0^*(2300)$ resonance from lattice QCD under chiral symmetry}

\author{Jing Luo\orcidlink{0009-0009-5148-1514}}\uestc 

\author{Bing Wu\orcidlink{0009-0004-8178-3015}}\email{wu.bing@uestc.edu.cn}\uestc 

\author{Pan-Pan Shi\orcidlink{0000-0003-2057-9884}}\ific 

\author{Meng-Lin Du\orcidlink{0000-0002-7504-3107}}\email{ du.ml@uestc.edu.cn}\uestc

\begin{abstract} 
We reanalyze the lattice spectra for $I=1/2$ $D\pi$ scattering in the $A_1^+$ irreducible representation from [\href{https://link.aps.org/doi/10.1103/PhysRevD.111.014503}{Phys. Rev. D 111, 014503 (2025)}] to investigate the impact of chiral and SU(3) flavor symmetries in $S$-wave $D\pi$ scattering and the $D_0^*(2300)$ resonance. By fitting the phase shifts obtained via L\"uscher's formula with both traditional and chirally modified effective-range expansion and $K$-matrix parameterizations, we find that the chiral factor shifts the extracted pole mass closer to the threshold (especially for resonances) and substantially reduces the resonance width. These findings are confirmed by unitarized chiral perturbation theory through a direct fit to the lattice spectra with both the single-channel and the $D\pi$-$D\eta$-$D_s\bar{K}$ coupled-channel schemes. Once the coupled channels are incorporated, the two-pole structure of the $D_0^*(2300)$ emerges. The trajectories of the two poles are investigated by varying the pion mass. 
\end{abstract}
\maketitle

\section{Introduction}
The discovery of the $D_{s0}^*(2317)$ by the BaBar collaboration in 2003~\cite{BaBar:2003oey} challenged the conventional quark model, in which mesons are interpreted as $\bar{q}q$ states, as it was significantly lighter than its conventional quark model expectation~\cite{Godfrey:1985xj,Godfrey:2015dva}. Since then, a growing number of resonant structures in the heavy-flavor sector have been observed that cannot be naturally accommodated within the traditional framework, and are therefore widely regarded as candidates for exotic states. Prominent examples include $\chi_{c1}(3872)$~\cite{Belle:2003nnu} (also known as $X(3872)$), $Z_c(3900)$~\cite{BESIII:2013ris,Belle:2013yex}, $Z_{cs}(3985)$~\cite{BESIII:2020qkh}, the pentaquark states $P_c(4312)$, $P_c(4440)$, and $P_c(4457)$~\cite{LHCb:2019kea}, as well as the doubly charmed tetraquark $T_{cc}^+(3875)$~\cite{LHCb:2021vvq,LHCb:2021auc}. These structures have attracted extensive theoretical and experimental investigation, for comprehensive reviews, see, \emph{e.g.}, Refs.~\cite{Chen:2016qju,Esposito:2016noz,Guo:2017jvc,Olsen:2017bmm,Liu:2019zoy,Brambilla:2019esw,Guo:2019twa,Albaladejo:2020tzt,Dong:2021bvy,Dong:2021juy,Chen:2022asf,ParticleDataGroup:2024cfk,Liu:2024uxn}. The discrepancy between the observed masses of $D_{s0}^*(2317)$ and $D_{s1}(2460)$~\cite{CLEO:2003ggt} and the predictions of the quark model~\cite{Godfrey:1985xj,Godfrey:2015dva} has motivated various interpretations, including hadronic molecules~\cite{Barnes:2003dj,vanBeveren:2003kd,Szczepaniak:2003vy,Kolomeitsev:2003ac,Chen:2004dy,Guo:2006fu,Guo:2006rp,Gamermann:2006nm,Yang:2021tvc}, tetraquark states~\cite{Cheng:2003kg,Maiani:2004vq}, and mixtures of $c\bar{q}$ with tetraquarks~\cite{Browder:2003fk}. The hadronic molecular picture is particularly appealing, as the mass difference between $D_{s1}(2460)$ and $D_{s0}^*(2317)$ is almost identical to that between $D^*$ and $D$, a natural consequence of heavy quark spin symmetry in the $D^{(*)}K$ interaction~\cite{Du:2017zvv}. In 2004, their nonstrange SU(3) partners, $D_{0}^*(2300)$~\cite{Belle:2003nsh,FOCUS:2003gru} (previously referred to as $D_0^*(2400)$) and $D_1(2430)$~\cite{ Belle:2003nsh}, were observed. This observation raises a puzzle: the two nonstrange mesons, \dast~and $D_1(2430)$, have masses almost equal to those of their strange counterparts, $D_{s0}^*(2317)$ and $D_{s1}(2460)$. 

The properties of the \dast~resonance can be studied through the $D\pi$ scattering. The first lattice Quantum chromodynamics (QCD) exploration of the \dast~and $D_1(2430)$ employing both $c\bar{q}$-type and $D\pi$-type interpolators at $M_\pi\approx 266$ MeV was presented in Ref.~\cite{Mohler:2012na}, yielding masses consistent with the experimental values. Subsequently, the Hadron Spectrum Collaboration (HSC) performed a $D\pi$–$D\eta$–$D_s\bar{K}$ coupled-channel simulation at $M_\pi \approx 391$ MeV. Using a $K$-matrix parameterization of the scattering amplitude, they found a pole corresponding to a $J^P=0^+$ bound state identified as the \dast. The HSC also computed the $I=1/2$ $D\pi$ scattering amplitude at a lighter pion mass, $M_\pi\approx 239$ MeV, in Ref.~\cite{Gayer:2021xzv}. The resulting amplitude exhibits a pole with a mass of $2200$ MeV and a width of $400$ MeV, which lies well below that of the $D_{s0}^*(2317)$ at the same pion mass~\cite{Cheung:2020mql}.

The amplitude parameterizations commonly used in experimental and lattice QCD analyses to extract the properties of the $D_0^*(2300)$, such as the Breit–Wigner (BW) formula, effective-range expansion (ERE), or $K$-matrix, do not respect chiral symmetry. As a consequence of the Goldstone theorem, chiral symmetry in QCD requires an energy-dependent pionic coupling, leading to
a mass smaller than their nominal values obtained without considering the chiral symmetry~\cite{Du:2019oki}. $D\pi$ scattering that respects chiral symmetry can be systematically described using heavy-meson chiral perturbation theory (ChPT)~\cite{Burdman:1992gh,Wise:1992hn,Yan:1992gz,Du:2016ntw,Du:2016xbh} in conjunction with unitarization methods~\cite{Oller:2000fj,Yao:2015qia,Guo:2015dha,Du:2017ttu}. In the absence of experimental data on the interactions between $D_{(s)}$ mesons and the Goldstone bosons $\phi=(\pi, K,\eta)$, the low-energy constants (LECs) in ChPT are typically determined via phenomenological approaches (\emph{e.g.}, Ref.~\cite{Du:2016tgp}) or lattice QCD simulations. The leading-order (LO) $D\phi$ scattering amplitude is fully determined by chiral symmetry and is known as the Weinberg–Tomozawa term. In Ref.~\cite{Liu:2012zya}, the scattering of pseudoscalar mesons ($\pi$, $K$) off charmed mesons ($D_{(s)}$) was studied in full lattice QCD using an extended L\"uscher formula. The scattering lengths for the $I=3/2$ $D\pi$, $D_s\pi$, $D_sK$, $I=0$ $D\bar{K}$, and $I=1$ $D\bar{K}$ channels were calculated at different pion masses. From these, the subtraction constant in the loop function and the ChPT LECs up to next-to-leading order (NLO) were determined. The resulting unitarized NLO chiral amplitudes support the $DK$ molecular interpretation of the $D_{s0}^*(2317)$. Using the parameters determined in that work, Ref.~\cite{Albaladejo:2016lbb} predicted the finite-volume energy levels for the $I = 1/2$ $D\pi$–$D\eta$–$D_s\bar{K}$ coupled-channel system, finding remarkable agreement with the lattice QCD calculations reported in Ref.~\cite{Moir:2016srx}. The successful description of the lattice data provides strong evidence that the chiral amplitudes in Refs.~\cite{Liu:2012zya,Albaladejo:2016lbb} are firmly grounded in QCD. It is worth noting that two $I(J^P) = 1/2(0^+)$ poles were found in the $D_0^*(2300)$ region at physical quark masses within the $D\pi$–$D\eta$–$D_s\bar{K}$ coupled-channel framework~\cite{Albaladejo:2016lbb}, located at $\left(2105^{+6}_{-8}-i~102_{-12}^{+10}\right)$ MeV and $\left(2451_{-26}^{+36}-i~134_{-8}^{+7}\right)$ MeV. These two poles couple predominantly to the $D\pi$ and $D_s\bar{K}$ channels, respectively. The lower pole belongs to the same SU(3) multiplet as the $D_{s0}^*(2317)$, thereby resolving the puzzle that the \dast~is almost as heavy as, or even heavier than, its strange partner. This pattern of two $I = 1/2$ states emerges naturally in chiral approaches respecting SU(3) flavor symmetry and has been identified in numerous studies (see, \emph{e.g.}, Refs.~\cite{Kolomeitsev:2003ac,Guo:2006fu,Wang:2012bu,Guo:2015dha,Guo:2018kno,Guo:2018tjx}). Further support for the two-pole structure of the \dast~comes from the analysis of high-quality LHCb data on $B^-\to D^+\pi^-\pi^-$ \cite{Aaij:2016fma} performed in Ref.~\cite{Du:2017zvv}, as well as analyses of $B_s^0\to\bar{D}^0K^-\pi^+$~\cite{Aaij:2014baa}, $B^0\to \bar{D}^0 \pi^-\pi^+$~\cite{Aaij:2015sqa}, $B^-\to D^+\pi^- K^-$~\cite{Aaij:2015vea}, and $B^0\to\bar{D}^0\pi^- K^+$~\cite{Aaij:2015kqa} in Ref.~\cite{Du:2019oki}. Moreover, Ref.~\cite{Du:2020pui} extracted the low-energy $S$-wave $D\pi$ phase shift for $B^-\to D^+\pi^-\pi^-$ from the data in a model-independent manner and found it to be consistent with the two-pole scenario, in contrast to the standard BW parameterization.

Recently, the ALICE Collaboration measured for the first time two-particle momentum correlation functions between charmed mesons ($D^{(*)\pm}$) and charged pions and kaons~\cite{ALICE:2024bhk}. The femtoscopy method is then used to extract the $D\pi$ scattering lengths by fitting the experimental correlation functions with a model that employs a Gaussian potential. The extracted values are found to be small and compatible with zero, in significant disagreement with predictions from lattice QCD and chiral effective theories. In Ref.~\cite{Du:2025beb}, a direct measurement of the $I=1/2$ $D\pi$ scattering length is proposed, which would not only provide valuable information about the \dast~but also help assess the reliability of scattering lengths extracted from femtoscopic studies.\footnote{For recent comments on the femtoscopy method, see Refs.~\cite{Epelbaum:2025aan,Molina:2025lzw}.} Furthermore, that work also proposes modified ERE and $K$-matrix parameterizations to restore chiral symmetry. Recently, Ref.~\cite{Yan:2024yuq} studied the $S$- and $P$-wave $I=1/2$ $D\pi$ scattering phase shifts using L\"uscher's formula on a series of $N_f = 2 + 1$ Wilson-Clover ensembles with pion masses $M_\pi \approx 133$, $208$, $305$, and $317$ MeV. The \dast~is found to be a virtual state at $M_\pi\approx 305$ and 317 MeV, while it becomes a resonance at $M_\pi\approx 133$ and 208 MeV. The pole at the physical pion mass is consistent with experimental results~\cite{Yan:2024yuq}, though no conclusive statement can be made due to the large uncertainty. We emphasize that the amplitude parameterizations employed in that work do not respect chiral symmetry and therefore lead to larger pole masses. Moreover, only one pole (instead of two) near the threshold region is found in Ref.~\cite{Yan:2024yuq} due to the omission of coupled-channel effects.\footnote{Although the two-hadron operators considered are limited to $D^{(*)}\pi$, coupled-channel effects ($D\pi$-$D\eta$-$D_s\bar{K}$) may arise from the inclusion of $N_f = 2 + 1$ dynamical quark flavors in Ref.~\cite{Yan:2024yuq}.} A systematic analysis of the $D_0^*(2300)$ using the lattice spectra from Ref.~\cite{Yan:2024yuq} that accounts for both chiral symmetry and SU(3) flavor symmetry, \emph{i.e.}, coupled-channel effects from $D\eta$ and $D_s\bar{K}$, calls for the unitarized ChPT.

In this work, we reanalyze $S$-wave $D\pi$ scattering and the $D_0^*(2300)$ resonance using the recent lattice QCD calculation on various ensembles for $I=1/2$ $D\pi$ scattering in Ref.~\cite{Yan:2024yuq}, systematically incorporating chiral symmetry and SU(3) flavor symmetry. In Sec.~\ref{sec:ere}, we parameterize the $S$-wave $I=1/2$ $D\pi$ scattering amplitude using the modified ERE and $K$-matrix approaches, which are constructed by including an energy-dependent factor to preserve the correct chiral behavior as proposed in Ref.~\cite{Du:2025beb}. After fitting the phase shifts extracted from the lattice spectra via L\"uscher's formula, we compare the results obtained from the modified parameterizations with those from the traditional ones to illustrate the effects of chiral symmetry. In Sec.~\ref{sec:UChPT}, using unitarized ChPT, we employ both the single-channel and the $D\pi$–$D\eta$–$D_s\bar{K}$ coupled-channel schemes to directly reanalyze the lattice spectra, thereby demonstrating the combined effects of chiral and SU(3) flavor symmetries. We conclude with a brief summary in Sec.~\ref{sec:summary}.

\section{Scattering analysis with modified ERE and $K$-matrix approaches}\label{sec:ere}
L\"uscher's quantization condition~\cite{Luscher:1985dn,Luscher:1986pf,Luscher:1990ux} relates the discrete energy spectra in a finite volume to the infinite-volume scattering phase shift~\cite{Briceno:2017max}. The finite cubic lattice breaks the rotational invariance and induces mixing among different partial waves. In the present work, we are interested only in $S$-wave $D\pi$ scattering and the properties of the \dast~resonance. To avoid uncertainties arising from $P$-wave and higher partial waves, we focus exclusively on the finite-volume spectra for the $A_1^+$ irreducible representation (irrep) in the rest frame, where contributions from partial waves with $\ell \geq 4$ can be safely neglected in the low-energy region. The spectra are taken from the various ensembles of Ref.~\cite{Yan:2024yuq} for $I=1/2$ $D\pi$ scattering. Consequently, for $S$-wave $D\pi$ scattering in relativistic kinematics, the unitarized amplitude takes the form
\begin{align}
    T(s)=\frac{8\pi\sqrt{s}}{k\cot\delta-ik}\ ,\label{unitaryofT}
\end{align}
where the $S$-wave scattering phase shift $\delta$ can be extracted via L\"uscher's formula: 
\bea\label{eq:Luescher}
k\cot\delta(k)=\frac{2}{\sqrt{\pi}L}\mathcal{Z}_{00}(1;\left(kL/2\pi\right)^2)\ ,
\eea
where $s$ is the center-of-mass (CM) energy squared, and $k = \sqrt{\lambda(s, M_D^2, M_\pi^2)}/(2\sqrt{s})$ is the magnitude of the three-momentum of either particle in the CM frame, and $\lambda(a,b,c)=a^2+b^2+c^2-2ab-2bc-2ca$ is the K\"all\'en function. The length of the cubic box in lattice simulations is denoted by $L$, and $\mathcal{Z}_{00}$ is the L\"uscher zeta function in the rest frame. The scattering momentum $k$ is related to the lattice scattering energy by
\begin{align}
    E(k) = \sqrt{M_D^2 + Z_D k^2} + \sqrt{M_\pi^2 + Z_\pi k^2}\ , \label{equ:lattice_energy}
\end{align}
where $Z_X$ ($X = \pi, D$) is the square of the speed of light on the lattice. $Z_X$ deviates from unity due to lattice discretization artifacts. However, $Z_X$ (particularly $Z_\pi$) is found to be close to unity, and the difference between observables obtained with the lattice-measured $Z_X$ and those with $Z_X = 1$ is small \cite{Yan:2024yuq}. Therefore, in this work, we simply set $Z_X = 1$.

\subsection{Traditional ERE and $K$-matrix parameterizations}
From the discrete energy levels in a finite volume, one can directly extract the phase shift at the corresponding energies via L\"uscher's formula, \emph{i.e.}, Eq.~(\ref{eq:Luescher}). To obtain the amplitude of Eq.~\eqref{unitaryofT} defined in the complex energy plane, one must resort to specific parameterizations of the $k\cot\delta$ term. A widely used parameterization for near-threshold phenomena is the ERE. For an $S$-wave scattering, the conventional ERE reads
\begin{align}
    k\cot\delta = \frac{1}{a_0} + \frac{1}{2}r_0 k^2 + \mathcal{O}(k^4)\ ,\label{eq:ERE}
\end{align}
where $a_0$ is the scattering length and $r_0$ the effective range. Another commonly used parameterization is the so-called $K$-matrix method:
\begin{align}
    T_K(s) = \frac{1}{K(s)^{-1} - i\rho(s)}\label{eq:TK}
\end{align}
with $\rho = k/(8\pi\sqrt{s})$ being the two-body phase space factor. The $K$-matrix can be parameterized as
\begin{align}
    K(s) = \frac{g^2}{m^2 - s} + \sum_n \gamma^{(n)}s^n\label{eq:K1}
\end{align}
or alternatively as
\begin{align}
    K(s) = \frac{1}{\sum\limits_{n=0} c^{(n)}s^n}\ ,\label{eq:K2}
\end{align}
where $g$, $\gamma^{(n)}$, and $c^{(n)}$ are undetermined parameters. 

\subsection{Chiral symmetry and modified parameterizations}
Neither the ERE in Eq.~\eqref{eq:ERE} nor the $K$-matrix approach in Eqs.~\eqref{eq:TK}–\eqref{eq:K2} respects chiral symmetry. As the (approximate) chiral symmetry of QCD is spontaneously broken to its diagonal subgroup ${\rm SU(3)}_V$, the lightest pseudoscalars $\pi$, $K$, and $\eta$ emerge as (pseudo-)Goldstone bosons. The low-energy strong interactions of these Goldstone bosons with hadrons are of derivative form and thus energy-dependent. In the chiral limit, \emph{i.e.}, $E_\pi \to 0$, the Goldstone bosons decouple from the hadrons in strong interactions. Consequently, the $D\pi$ scattering amplitude tends to vanish, exhibiting an Adler zero at the chiral limit $E_\pi \to 0$. The ERE and $K$-matrix parameterizations with correct chiral behavior near threshold were recently proposed in Ref.~\cite{Du:2025beb}. In this section, we first consider the traditional ERE and $K$-matrix parameterizations and then compare them with their chirally modified versions.

The $D\pi$ scattering amplitude, and consequently the low-energy parameters, depend on the pion mass. The lattice simulations in Ref.~\cite{Yan:2024yuq} for $I=1/2$ $D\pi$ scattering were performed on six ensembles with four different pion masses, specifically $M_\pi \approx 133$, $208$, $305$, and $317$ MeV. Among the six ensembles, F32P21 (F32P30) and F48P21 (F48P30) share the same lattice spacing and nearly the same pion mass of $208$ ($305$) MeV but differ in volume. Following Ref.~\cite{Yan:2024yuq}, we consider only the two lowest energy levels for the $A_1^+$ irrep on each ensemble. Consequently, for the ensembles H48P32 and C48P14, corresponding to $M_\pi \approx 317$ and $133$ MeV, respectively, the available lattice data constrain the amplitude to at most two free parameters. For $M_\pi \approx 305$ MeV and $208$ MeV, higher-order parameterizations can be employed, allowing for up to four free parameters. However, we find that higher-order parameterizations do not qualitatively alter the results, and thus we restrict ourselves to two-parameter parameterizations. Specifically, for the traditional ERE and $K$-matrix, we employ
\begin{align}
    k\cot\delta = \frac{1}{a_0} + \frac{1}{2}r_0 k^2\ ,\quad K(s) = \frac{g_0^2}{m_0^2 - s}\ .\label{eq:EREuse}
\end{align}
For the chirally modified ERE (MERE) and $K$-matrix (MK), we introduce an energy-dependent factor as
\begin{align}
    k\cot\delta =\frac{M_\pi}{E_\pi}\left(\frac{1}{a_0} + \frac{1}{2}r_0 k^2\right),\quad K(s) = \frac{E_\pi}{M_\pi} \frac{g_0^2}{m_0^2 - s}\ ,\label{eq:MEREuse}
\end{align}
 where
 \begin{align}
     E_\pi =\frac{s + M_\pi^2 - M_D^2}{2\sqrt{s}}\ .
 \end{align}
The factor $M_\pi$ ensures that the parameters in the traditional and modified versions have the same dimensions. In both cases, $a_0$ represents the scattering length defined at threshold.

\subsection{Traditional vs. chirally modified parameterizations}
\begin{figure}[tb]
\begin{center}
\includegraphics[height=0.36\linewidth]{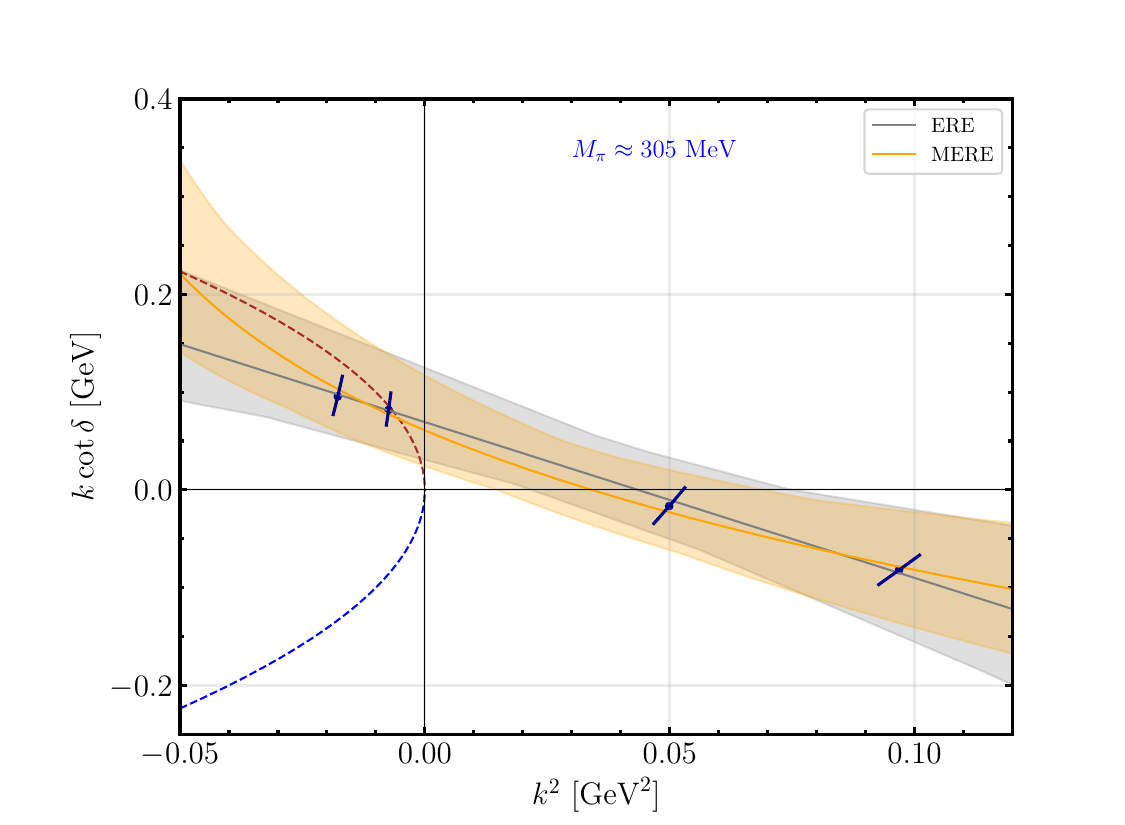}
\includegraphics[height=0.36\linewidth]{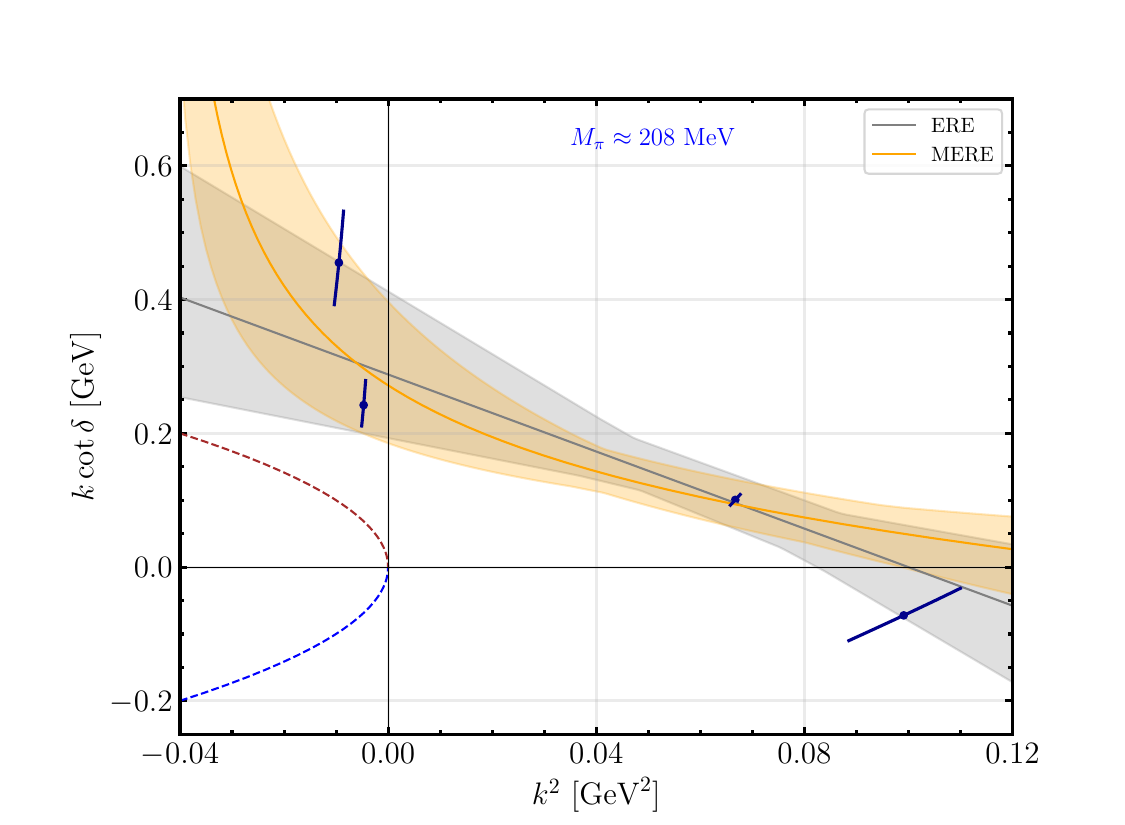}
\caption{Fitted $k\cot\delta$ for $I=1/2$ $D\pi$ scattering from Ref.~\cite{Yan:2024yuq} at $M_\pi \approx 305$ MeV (left panel) and $208$ MeV (right panel). The gray solid line and gray band represent the best fit and its $1\sigma$ uncertainty using the ERE formula in Eq.~\eqref{eq:EREuse}, while the orange solid line and orange band show the corresponding results using the MERE formula in Eq.~\eqref{eq:MEREuse}. The brown and blue dashed curves are the functions $-ik$ and $ik$, respectively. The dark blue arc segments denote the phase shift points extracted via L\"uscher's formula~\eqref{eq:Luescher} from the finite-volume spectra for the $A_1^+$ irrep in Ref.~\cite{Yan:2024yuq}, which are used in the fitting.}
\label{fig:ere}
\end{center}
\end{figure}
In Fig.~\ref{fig:ere}, we show the phase shifts extracted via L\"uscher's formula~\eqref{eq:Luescher} from the lattice energy spectra for the $A_1^+$ irrep of Ref.~\cite{Yan:2024yuq}, along with the fitting results using the ERE~\eqref{eq:EREuse} and MERE~\eqref{eq:MEREuse} parameterizations for $M_\pi \approx 305$ MeV and $208$ MeV, respectively. The extracted parameters are collected in Table~\ref{tab:parameters}.\footnote{It is noticed that the relative uncertainties of the parameters at $M_\pi=133$ MeV are substantially larger than those at the heavier pion masses. In particular, the uncertainties of \(g_0^2\) and \(r_0\) exceed their respective central values; these two parameters are connected by the approximate relation \(g_0^2\propto 1/r_0\) that follows from Eqs.~(8, 9). } Notably, the MERE curves deviate from the straight lines of the ERE due to the chiral factor $E_\pi$ in Eq.~\eqref{eq:MEREuse}. In particular, the phase shift $k\cot\delta$ exhibits a pole at $k^2 = -M_\pi^2$ (\emph{i.e.}, $E_\pi = 0$), corresponding to the Adler zero. Consequently, the chiral factor shifts the MERE poles relative to their ERE counterparts. For $M_\pi\approx 305$ MeV, both the ERE and MERE $k\cot\delta$ curves intersect the function $-ik$ below threshold, indicating a virtual state whose pole position lies at the intersection point on the second Riemann sheet (RS). Owing to the chiral factor, the MERE pole mass is slightly larger than the ERE counterpart (see, \emph{e.g.}, the left panel of Fig.~\ref{fig:ere}). For $M_\pi \approx 208$ MeV, neither the ERE nor the MERE intersect the function $\pm ik$, indicating that no bound or virtual state is found (see, \emph{e.g.}, the right panel of Fig.~\ref{fig:ere}). However, a pole with an imaginary part above the $D\pi$ threshold is found on the second RS, corresponding to a resonance. The pole positions are listed in Table~\ref{tab:pole:ere}. It is evident from Table~\ref{tab:pole:ere} that, the chiral factor shifts the real part (mass) of the pole closer to the threshold (significantly for resonances), regardless of whether the pole corresponds to a virtual state below threshold or a resonance with its real part above threshold, and also significantly reduces the absolute value of the imaginary part (half-width) for the resonance. Specifically, the mass shift from ERE to MERE is approximately $100$ MeV for $M_\pi \approx 208$ MeV and approximately $200$ MeV for $M_\pi\approx 133$ MeV, while the corresponding width shift is around $100$ MeV and $400$ MeV, respectively. This is not surprising, as chiral symmetry is more significant at $M_\pi \approx 133$ MeV than at $M_\pi \approx208$ MeV, and thus its effects are more pronounced. 
\renewcommand{\arraystretch}{1}
\begin{table}[tb]
\caption{Extracted parameters for the $S$-wave $I = 1/2$ $D\pi$ system from fits to the scattering phase shifts obtained via L\"uscher's formula~\eqref{eq:Luescher} from the lattice spectra for the $A_1^+$ irrep of the various ensembles in Ref.~\cite{Yan:2024yuq}, using the traditional (a) and modified (b) parameterizations given in Eqs.~\eqref{eq:EREuse} and \eqref{eq:MEREuse}, respectively.}
\centering
\footnotesize
\begin{minipage}{0.8\textwidth}
\centering
\vspace{0.25\baselineskip}  
\text{(a) Traditional parameterizations}
\begin{tabular}{c|cccc}
\toprule[2pt]
\makebox[4em][c]{$M_\pi$/MeV} & \makebox[6em][c]{$a_0$/fm} & \makebox[6em][c]{$r_0$/fm} & \makebox[6em][c]{$m_0$/GeV} & \makebox[6em][c]{$g_0^2$/GeV$^2$} \\
\midrule[0.5pt]
133 & 0.57(36) & $-0.6(18)$ & 2.51(77)& 321(776) \\
208 & 0.685(69) & $-1.14(18)$ & 2.312(13) & 166(26) \\
305 & 2.91(53) & $-0.620(89)$ & 2.344(12) & 287(42) \\
317 & 3.53(84) & $-0.732(77)$ & 2.3457(92) & 246(26) \\
\bottomrule[2pt]
\end{tabular}
\end{minipage}

\vspace{1.0\baselineskip}  

\begin{minipage}{0.8\textwidth}

\centering

\text{(b) Modified parameterizations}
\begin{tabular}{c|cccc}
\toprule[2pt]
\makebox[4em][c]{$M_\pi$/MeV} & \makebox[6em][c]{$a_0$/fm} & \makebox[6em][c]{$r_0$/fm} & \makebox[6em][c]{$m_0$/GeV} & \makebox[6em][c]{$g_0^2$/GeV$^2$} \\
\midrule[0.5pt]
133 & 0.58(36) & 1.5(18) & 1.69(80) & $-195(287)$ \\
208 & 0.726(74) & $-0.72(18)$ & 2.394(44) & 252(65) \\
305 & 3.24(66) & $-0.71(10)$ & 2.33(12) & 255(39) \\
317 & 4.7(17) & $-0.849(86)$ & 2.3280(85) & 213(22) \\
\bottomrule[2pt]
\end{tabular}
\end{minipage}
\label{tab:parameters}
\end{table}

In addition, for $M_\pi \approx 317$ and $305$ MeV, the chirally modified approaches yield an additional virtual state, which corresponds to a spurious pole~\cite{Ebert:2021epn} located deeper than the one previously discussed. This is evident in the left panel of Fig.~\ref{fig:ere}, where the MERE $k\cot\delta$ curve intersects the function $-ik$ at two points, each corresponding to a pole. The deeper virtual states lie far from the physical region and outside the domain of applicability of the approaches. We therefore exclude them from the present analysis.

\renewcommand{\arraystretch}{1}
\begin{table}[tb]
\caption{Pole positions $E_{\text{pole}}$ on the second RS for the $S$-wave $I = 1/2$ $D\pi$ scattering amplitude, obtained by fitting the scattering phase shifts derived via L\"uscher's formula~\eqref{eq:Luescher} from the lattice spectra for the $A_1^+$ irrep of the various ensembles in Ref.~\cite{Yan:2024yuq}, using the traditional~\eqref{eq:EREuse} and modified~\eqref{eq:MEREuse} parameterizations. The second RS is defined by $8\pi\sqrt{s}/T^{\text{II}}(s) = k\cot\delta + ik$. $m_{D\pi}$ in the last column denotes the $D\pi$ threshold mass of the corresponding ensemble.}
\footnotesize
\begin{tabular}{c|cc|cc|c}
\toprule[2pt]
\multirow{2}*{\makebox[4.50em][c]{$M_\pi$/MeV}} & \multicolumn{2}{c|}{Traditional $E_{\rm pole}$ (MeV)} &\multicolumn{2}{c|}{Modified $E_{\rm pole}$ (MeV)} &\multirow{2}*{\makebox[4.5em][c]{$m_{D\pi}$/MeV}} \\

 & \makebox[9.3em][c]{ERE} & \makebox[9.3em][c]{$K$-matrix} & \makebox[9.3em][c]{ERE} &  \makebox[9.3em][c]{$K$-matrix} & \\
\midrule[0.5pt]
$133$ & $2267(197)-i364(325)$ & $2119(132)-i421(367)$ & $2008(77)-i149(109)$ & $1993(85)-i181(129)$& $2037$ \\
$208$ & $2233(18)-i 167(28)$ & $2209(25)-i174(22)$ & $2137(16)-i126(5)$& $2134(15)-i120(5)$&$2109$ \\
$305$ & $2259(7)$ & $2260(7)$ & $  2261(7) $ & $ 2261(7) $ &$2271$ \\
$317$ & $2289(5)$ & $2290(5)$
 &$ 2292(4) $ & $  2293(4) $ & $2296$\\
\bottomrule[2pt]
\end{tabular}
\label{tab:pole:ere}
\end{table}

Although with correct chiral behavior, the modified ERE and $K$-matrix approaches can only be applied in the near-threshold energy region before the coupled-channel effect becomes important, and they do not account for the dependence on the pion mass dictated by chiral symmetry. A theoretical framework that simultaneously respects chiral symmetry, unitarity, and coupled-channel dynamics is provided by unitarized ChPT, which will be employed in the next section.

\section{Scattering analysis with the unitarized Chiral amplitude}\label{sec:UChPT}
\subsection{Infinite-volume unitarized chiral amplitude}
As the Goldstone bosons arising from spontaneously broken chiral symmetry in QCD, the low-energy interactions of the light pseudoscalar mesons $\phi = (\pi, K, \eta)$ with ground-state hadrons can be systematically described using SU(3) chiral effective theory. In ChPT for ground-state charmed mesons, the external four-momenta and the masses of the Goldstone bosons are counted as $\mathcal{O}(p)$ at low energies, whereas the temporal components of the four-momenta of the charmed mesons $D_{(s)}$ are counted as $\mathcal{O}(1)$ due to their nonvanishing masses in the SU(3) chiral limit. The spatial components of the four-momenta of the $D_{(s)}$ mesons, however, are counted as $\mathcal{O}(p)$; consequently, the virtuality of the internal $D_{(s)}$ propagators scales as $\mathcal{O}(p)$. According to the power counting rule, the LO chiral Lagrangian describing the $D_{(s)}\phi$ interaction is of $\mathcal{O}(p)$ and reads~\cite{Burdman:1992gh,Wise:1992hn,Yan:1992gz}
\bea\label{eq:Lag}
\mathcal{L}^{(1)}_{D\phi} = \mathcal{D}_\mu D \mathcal{D}^\mu D^\dagger-{M}_0^2D
D^\dagger\ ,
\eea
where the charmed meson triplet $D=(D^0,D^+,D_s^+)$ forms a basic representation of the unbroken subgroup ${\rm SU}_V(3)$ of the chiral symmetry, and $M_0$ denotes the mass of the triplet in the chiral limit. The covariant derivative is defined as
\bea\label{eq:cov}
\mathcal{D}_\mu D=D(\overset{\leftarrow}{\partial_\mu}+\Gamma_\mu^\dagger)\ ,
\qquad \mathcal{D}_\mu D^\dagger=(\partial_\mu+\Gamma_\mu)D^\dagger\ ,
\eea
with 
\begin{align}
    \Gamma_\mu &= \frac{1}{2}\left(u^\dagger\partial_\mu u+u\partial_\mu
u^\dagger\right),\quad u^2 = e^{i\sqrt{2}\Phi/F_\pi}\ ,\notag\\
\Phi &=\begin{pmatrix}
   \frac{1}{\sqrt{2}}\pi^0 +\frac{1}{\sqrt{6}}\eta  & \pi^+ &K^+  \\
     \pi^- &  -\frac{1}{\sqrt{2}}\pi^0 +\frac{1}{\sqrt{6}}\eta&K^0 \\
     K^-&\bar{K}^0&-\frac{2}{\sqrt{6}}\eta
\end{pmatrix} ,
\end{align}
where $F_\pi$ is the pion decay constant in the chiral limit. Discussions on the extension of chiral Lagrangians to next-to-next-to-leading order (NNLO) can be found in Refs.~\cite{Geng:2010vw,Yao:2015qia,Du:2016tgp,Du:2017ttu,Lutz:2022enz}, and the generalization to U(3) explicitly including the $\eta'$ meson is carried out in Ref.~\cite{Guo:2015dha}. 

In this work, we consider only the LO Lagrangian in Eq.~\eqref{eq:Lag}, \emph{i.e.}, the Weinberg–Tomozawa term, because at this order there are no unknown low-energy constants other than the pion decay constant $F_\pi$, and the form of the scattering amplitude is completely determined by chiral symmetry. For the process $D_1(p_1)\phi_1(p_2)\to D_2(p_3)\phi_2(p_4)$, the chiral amplitude takes the form
\bea\label{eq:V:WT}
V^{D_1\phi_1\to D_2\phi_2}
(s,u)=\frac{\mathcal{C}_\text{LO}(s-u)}{4F_\pi^2}\ ,
\eea
where the Mandelstam variables are defined as $s=(p_1+p_2)^2$ and $u=(p_1-p_4)^2$. The coefficient $\mathcal{C}_\text{LO}$ is process-dependent. Its values for the processes considered in this work are listed in Table~\ref{tab:LO}, following \emph{e.g.},  Refs.~\cite{Guo:2009ct,Wang:2012bu,Du:2017ttu}.

\renewcommand{\arraystretch}{1}
\begin{table}[tb]
\caption{The coefficients $\mathcal{C}_{\mathrm{LO}}$ for the $D\phi\to D\phi$ scattering potential in Eq.~\eqref{eq:V:WT} with strangeness $S=0$ and isospin $I=1/2$.}
\centering
\footnotesize
\begin{tabular}{c|cccccc}
\toprule[2pt]
\makebox[8em][c]{$(S,I)=(0,1/2)$}& \makebox[5.2em][c]{$D\pi\to D\pi$} & \makebox[5.2em][c]{$D\pi\to D\eta$} & \makebox[5.5em][c]{$D\pi\to D_s\bar{K}$} & \makebox[5.2em][c]{$D\eta\to D\eta$} & \makebox[6em][c]{$D\eta\to D_s\bar{K}$} & \makebox[6em][c]{$D_s\bar{K}\to D_s\bar{K}$} \\
\midrule[0.5pt]
$\mathcal{C}_{\mathrm{LO}}$ & $-2$ & $0$ & $-\sqrt{6}/2$ & $0$ & $-\sqrt{6}/2$ & $-1$ \\
\bottomrule[2pt]
\end{tabular}
\label{tab:LO}
\end{table}

Resonances are identified as poles of the corresponding scattering amplitude. Such poles cannot be obtained from a chirally expanded amplitude at finite orders because such expansions break unitarity. One way to restore unitarity is to unitarize the chiral potentials, which amounts to a resummation of the $s$-channel potentials. Before unitarizing, we perform a partial-wave projection of Eq.~\eqref{eq:V:WT} onto the $S$-wave as as follows:
\bea\label{eq:pwd}
V_0^{D_1\phi_1\to D_2\phi_2}(s) = \frac12\int_{-1}^{+1}\text{d}\cos\theta~ V^{D_1\phi_1\to D_2\phi_2}(s,u)\ ,
\eea
where the subscript $0$ denotes the $S$-wave, $\theta$ is the scattering angle between the incoming and outgoing particles in the CM frame, and the Mandelstam variable $u$ is expressed as
\bea
u(s,\cos\theta) &=& M_{D_1}^2+M_{\phi_2}^2-\frac{(s+M_{D_1}^2-M_{\phi_1}^2)(s+M_{\phi_2}^2-M_{D_2}^2)}{2s} \nonumber \\
&& +\frac{\cos\theta}{2s}\sqrt{\lambda(s,M_{D_1}^2,M_{\phi_1}^2)\lambda(s,M_{D_2}^2,M_{\phi_2}^2)}\ . \nonumber
\eea
In what follows, we suppress the superscript $D_1\phi_1\to D_2\phi_2$ for brevity. We employ the unitarization scheme~\cite{Oller:1998zr,Oller:2000fj}
\bea\label{eq:unitarization}
T(s) = \left[ 1-V_0(s)\cdot G(s)\right]^{-1}\cdot V_0(s)\ ,
\eea
which is widely used in studies of $D\phi$ scattering~\cite{Guo:2018tjx,Guo:2009ct,Du:2017ttu,Wang:2012bu}. Here, the Green's function $G(s)$ encodes two-body unitarity and is basically the two-point loop function
\bea
G(s)=i\int
\frac{d^4q}{(2\pi)^4}\frac{1}{(q^2-m_1^2+i\epsilon)
\left[(q+P)^2-m_2^2+i\epsilon\right]}\ ,\label{eq:loopintegral}
\eea
where $s \equiv P^2$ and $m_i$ ($i = 1, 2$) denote the masses of the two particles in the loop. This function can be evaluated using dimensional regularization or a once-subtracted dispersion relation. The explicit expression for $G(s)$ is given by~\cite{Oller:1998zr}:
\bea
G(s)&=& \frac{1}{16\pi^2}\Big\{ a(\mu) + \ln \frac{m_1^2}{\mu^2}+\frac{s-m_1^2+m_2^2}{2s} \ln \frac{m_2^2}{m_1^2}\nonumber \\
& & +\frac{\sigma (s)}{2s}\Big[ \ln \big(\sigma (s)+s+m_1^2-m_2^2\big) - \ln \big( \sigma (s)-s-m_1^2+m_2^2\big) \nonumber \\
& & +\ln \big(\sigma(s)+s-m_1^2+m_2^2\big)-\ln \big( \sigma (s)-s+m_1^2-m_2^2\big) \Big]\Big\}, \label{eq:Gs}
\eea
where $\sigma (s)=\sqrt{\lambda(s,m_1^2,m_2^2)}$ and $a(\mu)$ is a subtraction constant, with $\mu$ the renormalization scale. The function $G(s)$ is independent of the regularization scale $\mu$, as the explicit $\mu$-dependence is compensated by the subtraction constant $a(\mu)$. For simplicity, $\mu$ is typically chosen to be $1$ GeV. 

In coupled-channel schemes, the unitarized amplitude in Eq.~\eqref{eq:unitarization} should be understood in matrix form. Taking the $I=1/2$ $D\pi$-$D\eta$-$D_s\bar{K}$ system, which we will analyze in Sec.~\ref{sec:coupled}, as an example, the amplitude is given explicitly by
\begin{align}
    \mathbb{T}(s)=\left[ \mathbb{I}-
    \left(\begin{array}{ccc}
        V_{0,11} & V_{0,12} & V_{0,13}\\
        V_{0,12} & V_{0,22} & V_{0,23}\\
        V_{0,13} & V_{0,23} & V_{0,33} 
    \end{array}\right)\left(\begin{array}{ccc}
        G_{1} & 0 & 0\\
        0 & G_{2} & 0\\
        0 & 0 & G_{3} 
    \end{array}\right)
    \right]^{-1}\cdot \left(\begin{array}{ccc}
        V_{0,11} & V_{0,12} & V_{0,13}\\
        V_{0,12} & V_{0,22} & V_{0,23}\\
        V_{0,13} & V_{0,23} & V_{0,33} 
    \end{array}\right),\label{equ:Tcp}
\end{align}
where $\mathbb{I}$ is the $3\times 3$ identity matrix, $V_{0,ij}$ is the $S$-wave chiral potential from Eq.~\eqref{eq:pwd} for the transition from channel $i$ to channel $j$, and $G_{i}$ denotes the two-point loop function~\eqref{eq:Gs} for channel $i$. The indices $i,j=1,2,3$ correspond to the $D\pi$, $D\eta$, and $D_s\bar{K}$ channels, respectively. 

\subsection{Finite-volume unitarized chiral amplitude}
The unitarized chiral amplitudes in Eq.~\eqref{eq:unitarization} can also describe the finite-volume spectra obtained from lattice simulations by incorporating finite-volume effects (see, \emph{e.g.}, Ref.~\cite{D_ring_2011}). Since we focus solely on the $A_1^+$ irrep, the following discussion on finite-volume effects is restricted to the rest frame. 
In a finite volume with periodic boundary condition, the three-momentum running in the loop integral of Eq.~\eqref{eq:loopintegral} can only take discrete values, namely $\vec{q} = 2\pi\vec{n}/L$ with $\vec{n}\in \mathbb{Z}^3$. The finite-volume effects are incorporated by replacing the loop function $G(s)$ with~\cite{D_ring_2011,Guo:2018tjx}:
\bea
\tilde{G}(s,L) = G(s) + \Delta G(s,L)\ ,\label{equ:finite_G}
\eea
where 
\bea
\Delta G(s,L) = \frac{1}{L^3}\sum_{\vec{n}}^{|\vec{q\,}|< \Lambda} I(\vec{q\,}) - \int^{|\vec{q\,}|<\Lambda} \frac{\text{d}^3\vec{q}}{(2\pi)^3}I(\vec{q\,})\ ,
\eea
with 
\bea
I(\vec{q\,}) = \frac{w_1+w_2}{2w_1w_2\left[s-(w_1+w_2)^2\right]}\ , \quad w_i=\sqrt{|\vec{q\,}|^2+m_i^2}\ .
\eea
The quantity $\Delta G$ is independent of the three-momentum cutoff $\Lambda$ due to the cancellation of the cutoff dependence between the two terms in the limit $L\to\infty$. It is related to the L\"uscher zeta function as
\begin{align}
    \lim_{\Lambda\to\infty} \Delta G(s,L) = - \dfrac{1}{4\pi^{3/2}\sqrt{s}L}\mathcal{Z}_{00}(1;(kL/2\pi)^2)+i\dfrac{k}{8\pi\sqrt{s}}
\end{align}
up to exponentially suppressed terms. The finite-volume corrections to the short-distance potential $V_0(s)$ are exponentially suppressed and can therefore be neglected.

Although different partial waves mix in a finite volume due to the breaking of rotational invariance, there is no mixing among partial waves with $\ell < 4$ for the $A_1^+$ irrep. Consequently, the finite-volume energy levels are given by the poles of the finite-volume $T$-matrix, $\tilde{T}(s,L)$~\cite{Doring:2012eu}. For the single channel, the finite-volume $T$-matrix is described as 
\bea
\tilde{T}(s,L)=\left[1-V_0(s)\cdot \tilde{G}(s,L)\right]^{-1}\cdot V_0(s)\ .\label{equ:finite_T}
\eea
The finite-volume energy levels correspond to the solutions of the equation 
\bea\label{eq:det}
\text{det}\left[1-V_0(s)\cdot \tilde{G}(s,L)\right]=0\ .\label{equ:det1-VG_finite}
\eea

We now employ Eq.~\eqref{equ:det1-VG_finite} to fit the discrete finite-volume spectra of the $A_1^+$ irrep obtained from the lattice simulation in Ref.~\cite{Yan:2024yuq} for $I=1/2$ $D\pi$ scattering. Since the two-hadron operators considered in Ref.~\cite{Yan:2024yuq} are limited to $D^{(*)}\pi$, we first perform a single-channel fit. That is, the chiral potential $V_0(s)$ and the loop function $\tilde{G}(s,L)$ in Eq.~\eqref{equ:det1-VG_finite} are taken purely for the $I = 1/2$ $D\pi$ channel. However, given that the $N_f = 2 + 1$ dynamical quark flavors employed in Ref.~\cite{Yan:2024yuq} may induce coupled-channel effects due to the sea-quark contribution,\footnote{Although interpolating operators for the $D\eta$ and $D_s\bar K$ channels are not included in the lattice simulation~\cite{Yan:2024yuq}, these channels still affect the energy levels below their thresholds through sea-quark contributions.} we also carry out a coupled-channel fit that includes the $D\eta$ and $D_s\bar{K}$ channels. Specifically, the determinant condition becomes
\begin{align}
    {\rm det}\left[\mathbb{I}-
    \left(\begin{array}{ccc}
        V_{0,11} & V_{0,12} & V_{0,13}\\
        V_{0,12} & V_{0,22} & V_{0,23}\\
        V_{0,13} & V_{0,23} & V_{0,33} 
    \end{array}\right)\left(\begin{array}{ccc}
        \tilde{G}_{1} & 0 & 0\\
        0 & \tilde{G}_{2} & 0\\
        0 & 0 & \tilde{G}_{3} 
    \end{array}\right)
    \right]=0\ ,\label{equ:det1-VG_finite_couple}
\end{align}
where $\tilde{G}_{i}$ denotes the finite-volume loop function from Eq.~\eqref{equ:finite_G} for channel $i$.

\subsection{Chiral extrapolation of the pion decay constant $F_\pi$ and subtraction constant $a(\mu)$}\label{sec:chiralextrapolation}
The free parameters of the unitarized amplitude $T(s)$ in Eq.~\eqref{eq:unitarization}, or of the finite-volume unitarized amplitude $\tilde{T}(s,L)$ in Eq.~\eqref{equ:finite_T}, are the pion decay constant $F_\pi$ appearing in the LO chiral potential in Eq.~\eqref{eq:V:WT} and the subtraction constant $a(\mu)$ entering the two-point loop function in Eq.~\eqref{eq:Gs}. Since the lattice simulations are performed at unphysical pion masses, we must account for the pion mass dependence of these constants, which, however, enters at NLO. Using the results from one-loop chiral perturbation theory~\cite{Gasser:1983yg}, the pion decay constant $F_\pi$ can be obtained via the chiral extrapolation formula~\cite{Baru:2013rta}: 
\bea\label{eq:Fpi}
F_\pi(M_\pi) = F_\pi^\text{ph}\left[1+\left(1-\frac{F_0}{F_\pi^\text{ph}}\right)(\xi^2-1)-\frac{(M_\pi^\text{ph})^2}{8\pi^2F_0^2}\xi^2\log \xi\right],
\eea
where $\xi = M_\pi/M_\pi^\text{ph}$, with $M_\pi^{\rm ph}=138$ MeV being the physical pion mass, $F_\pi^{\rm ph}=92.1$ MeV the physical pion decay constant, and $F_0 = F_\pi(M_\pi=0)$ the pion decay constant in the chiral limit, which is to be determined. Similarly, taking into account the pion mass dependence, the subtraction constant $a(\mu)$ can be parameterized as 
\bea\label{eq:amu}
a(\mu) = \bar{a}(\mu) + \beta M_\pi^2\ ,
\eea
where the two parameters $\bar{a}$ and $\beta$ are to be determined.

\subsection{Single-channel analysis}\label{sec:single}
\renewcommand{\arraystretch}{1}
\begin{table}[tb]
\caption{Meson masses and corresponding coefficients for the lattice ensembles used in Ref.~\cite{Yan:2024yuq}, with the kaon mass $M_K$ and the average up- and down-quark mass $\hat{m}$ taken from a previous study on the same ensembles in Ref.~\cite{CLQCD:2023sdb}.}
\centering
\footnotesize
\begin{tabular}{c|cccccc}
\toprule[2pt]
\makebox[6em][c]{ensemble} & \makebox[5em][c]{volume} & \makebox[6em][c]{$a/$fm} & \makebox[5em][c]{$M_\pi/$MeV} & \makebox[6em][c]{$M_D/$MeV} & \makebox[6em][c]{$M_K/$MeV} & \makebox[5em][c]{$\hat{m}/$MeV}  \\
\midrule[0.5pt]
C48P14 & $48^3\times 96$ & $0.10530(18)$ & $133.1(16)$ & $1903.73(34)$ & $510.0(10)$& $3.638(83)$  \\
F32P21 & $32^3\times$ 64 & 0.07746(18) & 206.8(21) & 1901.3(11) & 492.0(17) & $8.58(16)$\\
F48P21 & $48^3\times 96$ & 0.07746(18) & 208.12(70) & 1900.71(48) & 493.0(14)& $8.59(08)$\\
F32P30 & $32^3\times 96$ & 0.07746(18) & 305.81(71) & 1965.7(10) & 524.6(18)& $18.54(12)$\\
F48P30 & $48^3\times 96$ & 0.07746(18) & 304.98(50) & 1966.20(57) & 523.6(14)& $18.511(92)$\\
H48P32 & $48^3\times 144$ & 0.05187(26) & 317.00(68) & 1979.41(85) & 536.1(30)& $19.42(05)$\\
\bottomrule[2pt]
\end{tabular}
\label{tab:setting}
\end{table}
We first perform a single-channel fit to the discrete finite-volume spectra of the $A_1^+$ irrep for $I=1/2$ $D\pi$ scattering. Specifically, the energy levels on the six ensembles are used to determine the three parameters $F_0$, $\bar{a}(\mu)$, and $\beta$ in Eq.~\eqref{equ:det1-VG_finite}. For each ensemble, we utilize the first two energy levels in the $A_1^+$ irrep, consistent with the analysis in Ref.~\cite{Yan:2024yuq}.
The masses of the mesons used in the single-channel fit, namely $M_D$ and $M_\pi$, are taken from the lattice simulation and listed in Table~\ref{tab:setting}. The fit results are shown in Fig.~\ref{fig:single:fit}, and the corresponding fitted parameters are collected in Table~\ref{tab:paras:T}. Notably, the extracted pion decay constant in the chiral limit, $F_0$, is close to the value of $85$ MeV reported in Ref.~\cite{Becirevic:2012pf}. 
\begin{figure}[htb]
\centering
\setlength{\tabcolsep}{-1pt}  
\begin{tabular}{@{}cccccc@{}}
\includegraphics[height=0.37\textheight]{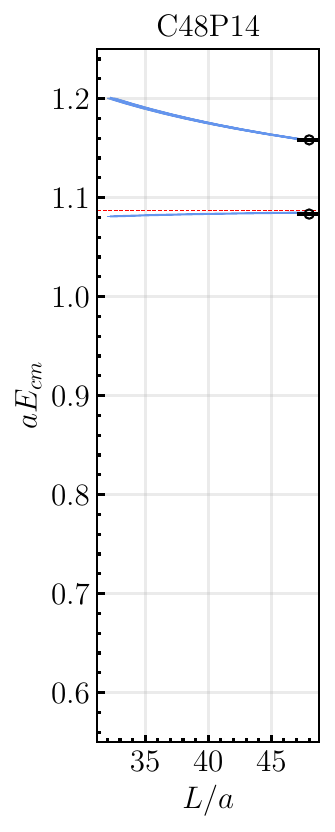} &  
\includegraphics[height=0.37\textheight]{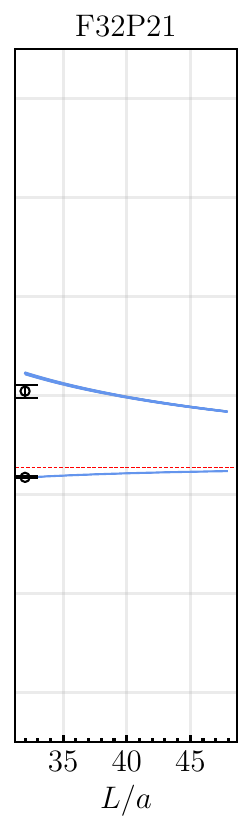} &
\includegraphics[height=0.37\textheight]{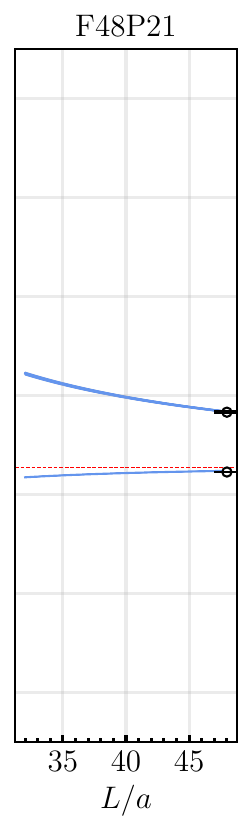} &
\includegraphics[height=0.37\textheight]{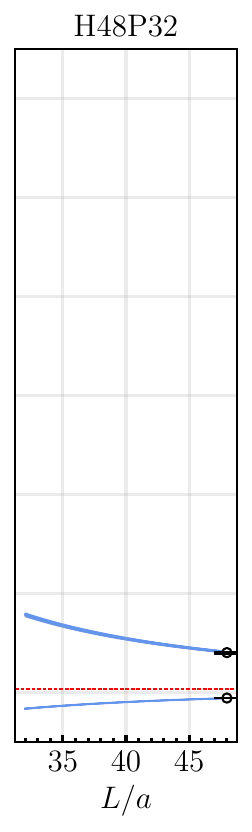} &
\includegraphics[height=0.37\textheight]{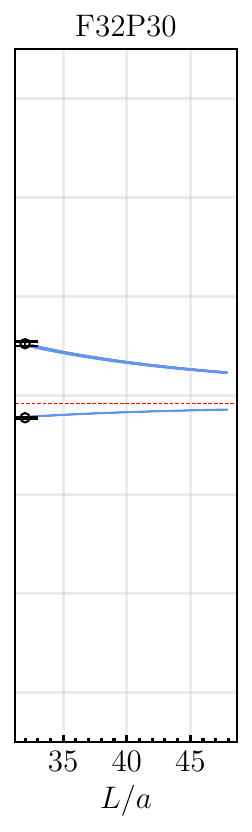} &
\includegraphics[height=0.37\textheight]{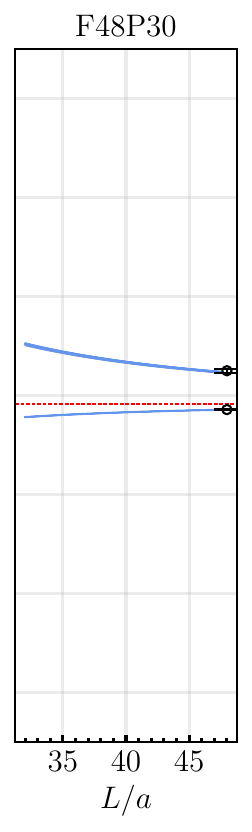}
\end{tabular}
\caption{Fitted finite-volume spectra for the $A_1^+$ irrep of the various ensembles in Ref.~\cite{Yan:2024yuq} for $I=1/2$ $D\pi$ scattering, obtained using the single-channel scheme in Eq.~\eqref{equ:det1-VG_finite}. The blue shaded areas indicate the $1\sigma$ error bands of the fit, the red dashed line marks the $D\pi$ threshold, and the black circles with error bars represent the lattice energy levels. Note that the 1$\sigma$ error band is extremely narrow and nearly invisible, as a result of the small uncertainties in the lattice data.}
\label{fig:single:fit}
\end{figure}
\renewcommand{\arraystretch}{1}
\begin{table}[tb]
\caption{Extracted parameters and $\chi^2/\text{d.o.f.}$ from fits to the finite-volume spectra for the $A_1^+$ irrep of the various ensembles in Ref.~\cite{Yan:2024yuq} for $I = 1/2$ $D\pi$ scattering, using the single-channel and coupled-channel schemes defined in Eqs.~\eqref{equ:det1-VG_finite} and~\eqref{equ:det1-VG_finite_couple}, respectively, with the renormalization scale set to $\mu = 1$ GeV. The corresponding fit results are shown in Figs.~\ref{fig:single:fit} and~\ref{fig:coupled:fit}, respectively.
}
\centering
\footnotesize
\begin{tabular}{c|cc}
\toprule[2pt]
\makebox[5em][c]{}& \makebox[8em][c]{single-channel} & \makebox[8em][c]{coupled-channel} \\
\midrule[0.5pt] 
$\bar a(\mu)$ & $-1.38(5)$ & $-1.49(4)$ \\
$F_0/\text{GeV}$ &  $0.0862(7)$  & $0.0886(7)$ \\
$\beta/\text{GeV}^{-2}$  & $-7.80(101)$ & $-3.98(70)$ \\
$\chi^2/ \text{d.o.f}$  & $2.35$ & $1.99$ \\
\bottomrule[2pt]
\end{tabular}
\label{tab:paras:T}
\end{table}

With the fitted parameters, we can search for poles of the scattering amplitude on various RSs. Bound states are identified as poles on the physical RS, whereas resonances and virtual states correspond to poles on unphysical RSs. Different RSs can be characterized by the sign of the imaginary part of the loop function $G(s)$ on the unitary cuts. Each $G_i(s)$ in Eq.~\eqref{equ:Tcp} has two sheets: the physical (first) RS and the unphysical (second) RS, denoted by $G_i(s)$ and $G_i^\text{II}(s)$, respectively. The expression for $G_i^\text{II}(s)$ can be obtained via analytic continuation~\cite{Oller:1998zr}:
\bea
G_i^\text{II}(s) = G_i(s)+2i\rho_i(s)\ ,\label{equ:G_II}
\eea
where $\rho_i(s)$ is the two-body phase space for channel $i$. Finally, we present in Table~\ref{tab:single-pole} the pole positions on the second RS with the different ensemble mass configurations, corresponding to different values of $M_\pi$. 

By comparing the poles of single-channel unitarized chiral amplitudes (Table~\ref{tab:single-pole}) with those extracted from phase-shift fits using traditional (ERE, $K$-matrix) and modified (MERE, MK) parameterizations (Table~\ref{tab:pole:ere}), we observe a clear consistency in the pole types across all three methods: resonances at $M_\pi \approx 133$ and $208$ MeV, and virtual states at $M_\pi \approx 305$ and $317$ MeV. However, sizable differences appear in the pole positions (especially for resonances). Furthermore, the pole positions from the single-channel unitarized chiral amplitude are consistent with those from the modified parameterizations for all pion masses. In comparison to the widely used ERE, the inclusion of chiral symmetry tends to move the pole masses closer to threshold (especially for resonances) and significantly reduces the absolute value of the resonance width. Specifically, the mass shift from ERE to the single-channel unitarized chiral amplitude is approximately 100 MeV at $M_\pi \approx 133$ MeV and about 70 MeV at $M_\pi \approx 208$ MeV. The positions of the virtual poles obtained from the modified parameterizations are in excellent agreement with those from the unitarized chiral amplitude, within the estimated uncertainties. 

Moreover, similar to the MERE and MK approaches, an extra deep virtual pole is found for $M_\pi\approx305$ or 317 MeV. We do not list such poles in Tables because they lie beyond the scope of the present framework. Their large distance (if they exist) from the physical region would render them unobservable in physical quantities due to screening by the near‑threshold pole. The appearance of such poles can be seen by the investigation of the pole trajectory below.

In addition, we investigate the trajectory of the pole position as a function of the pion mass. The dependence of $F_\pi$ and $a(\mu)$ on the pion mass is given in Eqs.~\eqref{eq:Fpi} and \eqref{eq:amu}, respectively. The pion mass dependence of the charmed meson mass can be obtained from ChPT for charmed mesons. Here we employ a simple parameterization~\cite{Liu:2012zya,Guo:2015dha}:
\bea
M_D^2 = \bar{M}_D^2+ \alpha_D M_\pi^2\ ,\label{equ:MD}
\eea
where $\bar{M}_D$ denotes the mass of the $D$ meson in the chiral limit. By fitting to the values of $M_D$ obtained from lattice simulations on various ensembles~\cite{Yan:2024yuq} (\emph{i.e.}, for different $M_\pi$), as listed in Table~\ref{tab:setting}, we obtain $\bar{M}_D=1868.6(6)\ \text{MeV}$ and $\alpha_D=3.911(34)$. With these results, we can construct the dependence of the pole position on the pion mass within the single-channel scheme. Specifically, we select 25 pion mass points ranging from 30 to 700 MeV and compute the corresponding pole trajectory. The resulting behavior is shown in Fig.~\ref{fig:single:trajectory}. For $M_\pi$ between 30 MeV and 300 MeV, the conjugate poles correspond to a resonance. As the pion mass increases, the ratio of the resonance mass to the $D\pi$ threshold gradually decreases, and the width becomes narrower. In the range $300~\text{MeV} < M_\pi < 325~\text{MeV}$, the poles turns to two virtual states. With further increase in $M_{\pi}$, the ratio of the pole mass to the $D\pi$ threshold decreases for one of the virtual states, while it increases for the other one and reaches unity at $M_{\pi}=325$ MeV. For $M_\pi > 325$ MeV, the latter pole moves onto the first RS and becomes a bound state. The ratios of the masses of the two poles (one virtual and one bound) to the $D\pi$ threshold decrease with the increase of pion mass. This behavior of the pole trajectory is consistent with that described in Ref.~\cite{Guo:2015dha}, which is based on the $D\pi$-$D\eta$-$D_s\bar{K}$ coupled-channel unitarized chiral amplitudes up to NLO.

\renewcommand{\arraystretch}{1.1}
\begin{table}[tb]
\caption{Pole positions $E_{\text{pole}}$ on the second RS for the $S$-wave $I = 1/2$ $D\pi$ unitarized chiral amplitude, obtained from fits to the finite-volume spectra for the $A_1^+$ irrep of various ensembles in Ref.~\cite{Yan:2024yuq} for $I=1/2$ $D\pi$ scattering using the single-channel unitarized chiral amplitude in Eq.~\eqref{equ:det1-VG_finite}. The corresponding fit results are shown in Fig.~\ref{fig:single:fit}. $m_{D\pi}$ denotes the value of the $D\pi$ threshold.}
\centering
\footnotesize
\begin{tabular}{c|c|c|c|c}
\toprule[2pt]
\makebox[6em][c]{ensemble}& \makebox[5em][c]{$M_\pi/$MeV} & \makebox[6em][c]{$M_D/$MeV} & \makebox[15em][c]{$E_{\rm pole}$ (MeV)} & \makebox[6em][c]{$m_{D\pi}/$MeV}\\
\midrule[0.5pt]
C48P14 &$133 $ & $1904 $ & $2144^{+1}_{-1}-i115^{+8}_{-8}$ & 2037 \\
F32P21/F48P21 &208 & 1901 & $2156^{+3}_{-2}-i115^{+6}_{-6}$ & 2109\\
F32P30/F48P30 & 305  & 1966 & $  2263^{+5}_{-13} $ & 2271\\
H48P32 & 317 & 1979 & $ 2295^{+1}_{-2} $ & 2296\\
\bottomrule[2pt]
\end{tabular}
\label{tab:single-pole}
\end{table}

\begin{figure}[tb]
\begin{center}
\begin{tikzpicture}
    \node[anchor=south west,inner sep=0] (image) at (0,0) {
        \includegraphics[width=0.75\linewidth]{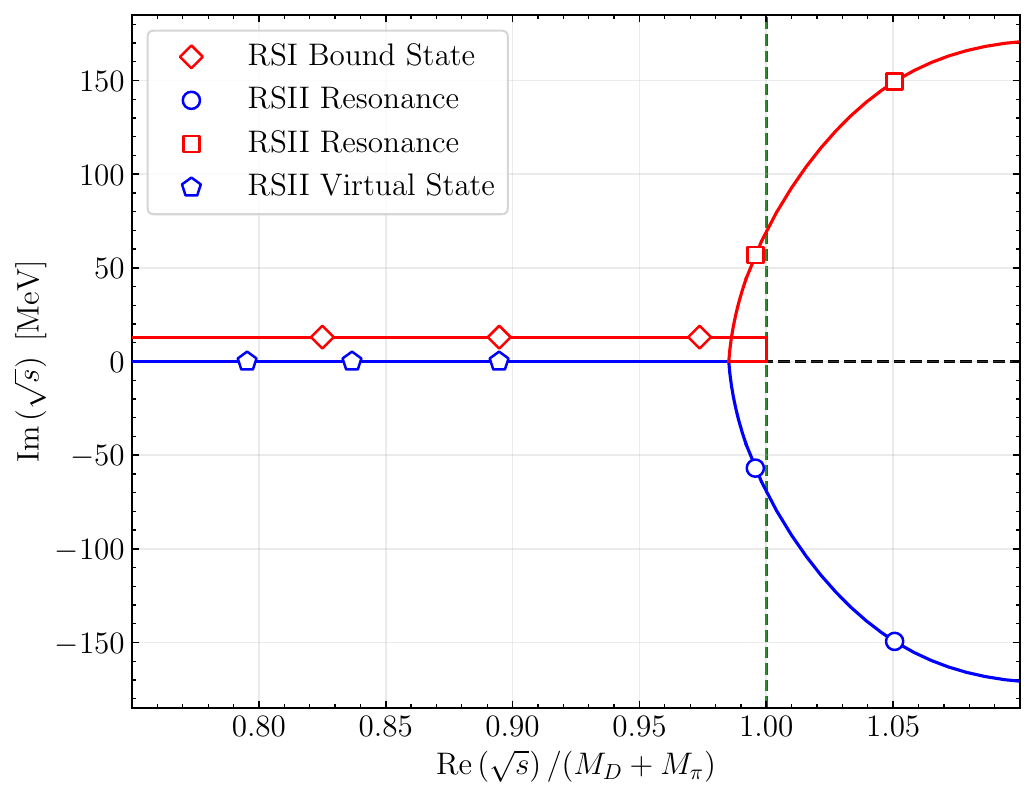}
    };

    \begin{scope}[x={(image.south east)},y={(image.north west)}]
 
       \draw[-{Stealth[open, length=8pt, width=6pt,fill=white]},thick,blue] (0.55,0.547) -- (0.54,0.547);
       \draw[-{Stealth[open, length=8pt, width=6pt,fill=white]},thick,blue] (0.365,0.547) -- (0.355,0.547);
       \draw[-{Stealth[open, length=8pt, width=6pt,fill=white]},thick,blue] (0.820,0.242) -- (0.804,0.264);

       \draw[-{Stealth[open, length=8pt, width=6pt,fill=white]},thick,red] (0.57,0.578) -- (0.56,0.578);
       \draw[-{Stealth[open, length=8pt, width=6pt,fill=white]},thick,red] (0.809,0.837) --(0.800,0.825);
       \draw[-{Stealth[open, length=8pt, width=6pt,fill=white]},thick,red] (0.385,0.578) -- (0.375,0.578);
       \draw[-{Stealth[open, length=8pt, width=6pt,fill=white]},thick,red] (0.725,0.547) -- (0.735,0.547);

       \draw[-{Stealth[open, length=8pt, width=6pt,fill=white]},thick,black] (0.775,0.471) -- (0.707,0.531);
       \draw[-{Stealth[open, length=8pt, width=6pt,fill=white]},thick,black] (0.816,0.602) -- (0.748,0.558);
       \draw[-{Stealth[open, length=8pt, width=6pt,fill=white]},thick,black] (0.682,0.776) -- (0.738,0.702);

        \node[rotate=90, green!50!black, font=\scriptsize] at (0.72,0.24) {$D\pi$ threshold};
        \node[blue, font=\scriptsize] at (0.871,0.170) {1};
        \node[blue, font=\scriptsize] at (0.718,0.394) {2};
        \node[blue, font=\scriptsize] at (0.468,0.520) {3};
        \node[blue, font=\scriptsize] at (0.325,0.520) {4};
        \node[blue, font=\scriptsize] at (0.223,0.520) {5};
        \node[red, font=\scriptsize] at (0.844,0.921) {1};
        \node[red, font=\scriptsize] at (0.707,0.700) {2};
        \node[red, font=\scriptsize] at (0.656,0.596) {3};
        \node[red, font=\scriptsize] at (0.466,0.596) {4};
        \node[red, font=\scriptsize] at (0.296,0.596) {5};
        \node[black, font=\scriptsize] at (0.653,0.792) {$M_{\pi} = 266$ MeV};
        \node[black, font=\scriptsize] at (0.833,0.452) {$M_{\pi} = 300$ MeV};
        \node[black, font=\scriptsize] at (0.839,0.626) {$M_{\pi} = 325$ MeV};
    \end{scope}
\end{tikzpicture}
\end{center}
\raggedright
\caption{Pole trajectory of the single-channel unitarized chiral amplitude for $I = 1/2$ $D\pi$ scattering around 2200 MeV, where $n$ is defined by $M_\pi = n M_\pi^{\text{ph}}$. See Eq.~\eqref{equ:G_II} for the definition of the RS.}
\label{fig:single:trajectory}
\end{figure}
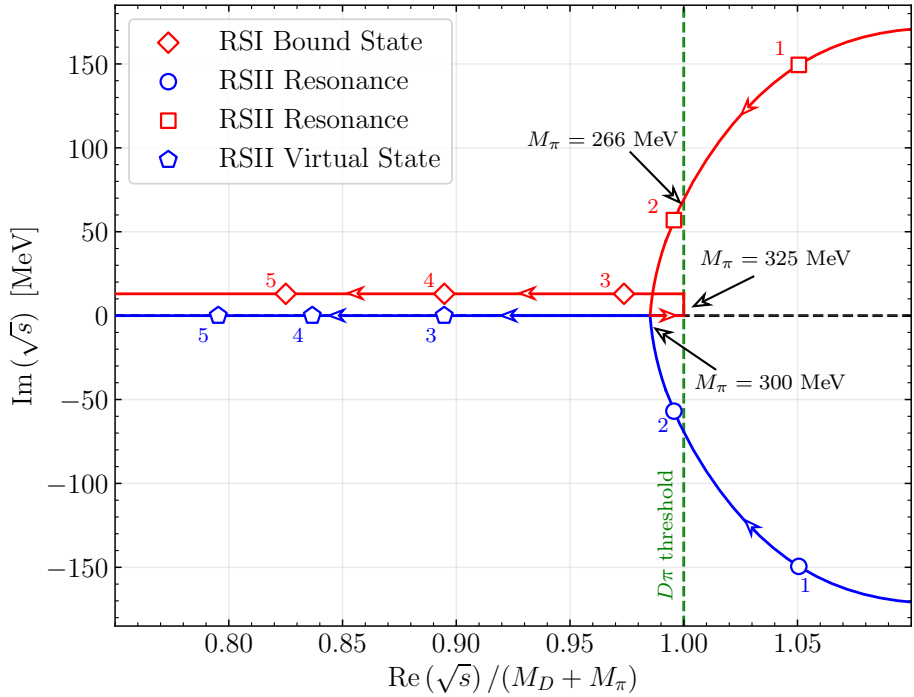

\subsection{Coupled-channel analysis}\label{sec:coupled}
We also perform a coupled-channel fit incorporating the $D\pi$, $D\eta$, and $D_s\bar{K}$ channels, as given in Eq.~\eqref{equ:det1-VG_finite_couple}, to the discrete finite-volume spectra of the $A_1^+$ irrep in Ref.~\cite{Yan:2024yuq} for $I=1/2$ $D\pi$ scattering. The masses of the $M_D$, $M_\pi$, and $M_{K}$ mesons used in the coupled-channel fit are taken from the lattice simulations~\cite{Yan:2024yuq,CLQCD:2023sdb} as listed in Table~\ref{tab:setting}. The mass $M_{\eta}$ is determined by the Gell-Mann–Okubo formula, and $M_{D_{s}}$ is approximately taken as the mass of the $D$ meson plus the mass difference between the strange quark and the light quark: 
\bea
 M_{\eta}=\sqrt{ \frac{4 M_K^2 - M_{\pi}^2} {3}}\ ,\quad M_{D_s}= M_D + \frac{M_K^2 - M_{\pi}^2}{B}\ ,\label{equ:Meta_MDs}
\eea
where $B$ is the average value over the various ensembles in Ref.~\cite{Yan:2024yuq}. For each ensemble, $B = M_\pi^2 / (2\hat{m})$, with the light quark mass $\hat{m}$ being the average mass of the up and down quarks, as provided in Ref.~\cite{CLQCD:2023sdb} and listed in the last column of Table~\ref{tab:setting}.

As mentioned in Sec.~\ref{sec:chiralextrapolation}, we employ a uniform subtraction constant $a(\mu)$ for the three-channel 
loop functions $G_i(s)$ in Eq.~\eqref{equ:det1-VG_finite_couple}. Consequently, as in the single-channel scheme described in Sec.~\ref{sec:single}, there are also three free parameters to be determined from the finite-volume spectra, namely $F_0$, $\bar{a}(\mu)$, and $\beta$. The fit results for the energy levels and the extracted parameters are shown in Fig.~\ref{fig:coupled:fit} and the third column of Table~\ref{tab:paras:T}, respectively. Notably, the coupled-channel fit yields noticeably better results than the single-channel approach. We attribute this improvement to the fact that the lattice data are derived from the symmetric \(N_f = 2 + 1\) Wilson-Clover ensembles in Ref.~\cite{Yan:2024yuq}, where the sea-quark contributions are effectively incorporated through their hadronization into the $D\eta$ and $D_s \bar K$ channels.  

\begin{figure}[tb]  
\centering
\setlength{\tabcolsep}{-1pt}  
\renewcommand{\arraystretch}{0}  
\begin{tabular}{@{}cccccc@{}}  
\includegraphics[height=0.368\textheight]{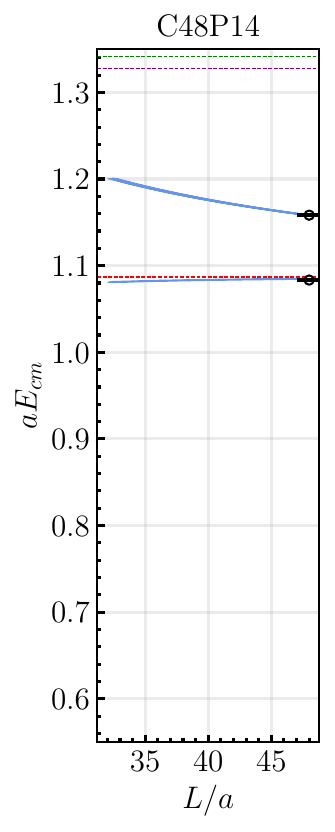} &
\includegraphics[height=0.368\textheight]{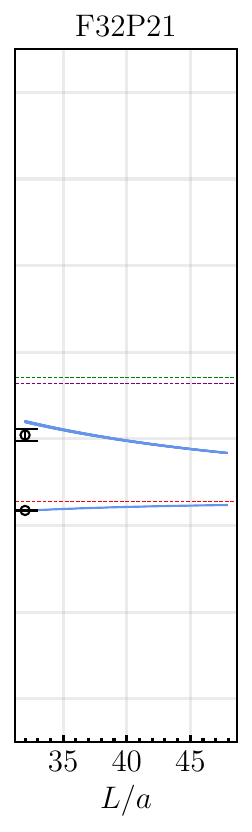} &
\includegraphics[height=0.368\textheight]{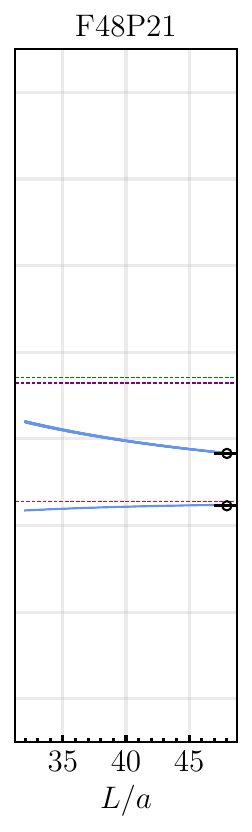} &
\includegraphics[height=0.368\textheight]{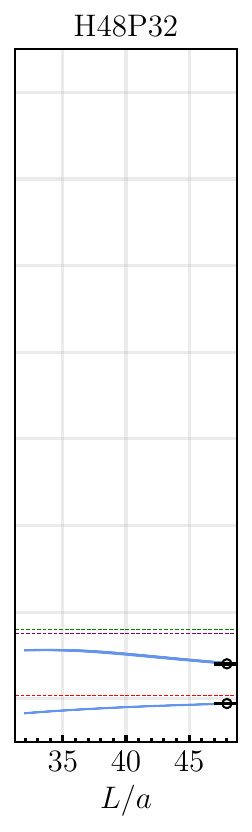} &
\includegraphics[height=0.368\textheight]{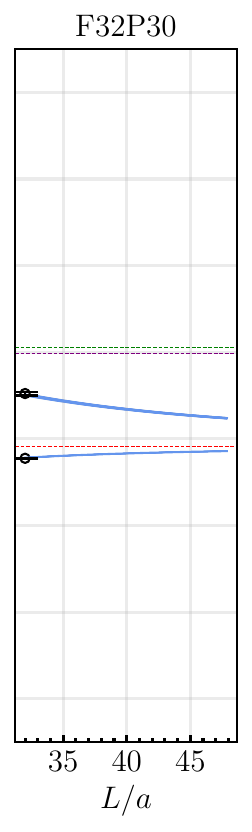} &
\includegraphics[height=0.368\textheight]{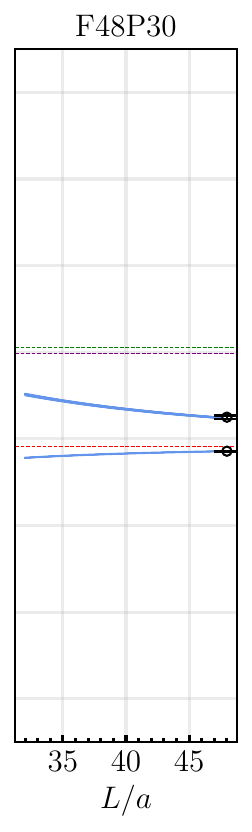}
\end{tabular}
\caption{Fitted finite-volume spectra for the $A_1^+$ irrep of the various ensembles in Ref.~\cite{Yan:2024yuq} for $I=1/2$ $D\pi$ scattering, obtained using the $D\pi$-$D\eta$-$D_s\bar{K}$ coupled-channel scheme in Eq.~\eqref{equ:det1-VG_finite}. The blue shaded areas indicate the $1\sigma$ error bands of the fit; the red, purple, and green dashed lines mark the $D\pi$, $D\eta$, and $D_s\bar{K}$ thresholds, respectively; and the black circles with error bars represent the lattice energy levels. Note that the 1$\sigma$ error band is extremely narrow and nearly invisible, as a result of the small uncertainties in the lattice data.}
\label{fig:coupled:fit}
\end{figure}

In the $I=1/2$ $D\pi$-$D\eta$-$D_s\bar{K}$ coupled-channel scheme, there exist $2^3 = 8$ RSs for each scattering matrix element. Different RSs can be accessed by appropriately choosing the loop functions $G_i(s)$ or $G_i^\text{II}(s)$. The RSs are denoted by $[\alpha_1 \alpha_2 \alpha_3]$, where $\alpha_i = 0$ represents the physical sheet (associated with $G_i(s)$) and $\alpha_i = 1$ represents the unphysical sheet (associated with $G_i^\text{II}(s)$) for channel $i$, as defined in Eq.~\eqref{equ:G_II}. In this convention, $[000]$ denotes the physical sheet, while $[100]$, $[110]$, $[111]$, and so on correspond to the second, third, fourth, etc., sheets, respectively.

\renewcommand{\arraystretch}{1}
\begin{table}[tb]
\caption{Pole positions $E_{\text{pole}}$ on the second ([100]) and third ([110]) RSs for the $S$-wave $I = 1/2$ $D\pi$ unitarized chiral amplitude, obtained from fits to the finite-volume spectra for the $A_1^+$ irrep of various ensembles in Ref.~\cite{Yan:2024yuq} for $I=1/2$ $D\pi$ scattering using the $D\pi$–$D\eta$–$D_s\bar{K}$ coupled-channel unitarized chiral amplitude in Eq.~\eqref{equ:det1-VG_finite_couple}. The corresponding fit results are shown in Fig.~\ref{fig:coupled:fit}. $m_{D\pi}$, $m_{D\eta}$, and $m_{D_s\bar{K}}$ denote the values of the thresholds of the respective channels. All values are given in units of MeV.}
\centering
\footnotesize  
\begin{tabular}{c|c|cc|ccc}
\toprule[2pt]
\makebox[4.5em][c]{\text{ensemble}} & \makebox[3em][c]{$M_\pi$} & \makebox[11em][c]{$E_{\rm pole}$ in $[100]$} & \makebox[11em][c]{$E_{\rm pole}$ in $[110]$} &  \makebox[3em][c]{$m_{D\pi}$}& \makebox[3em][c]{$m_{D\eta}$} & \makebox[3em][c]{$m_{D_s\bar{K}}$}\\ 
\midrule[0.5pt] 
C48P14 & 133 & $2149^{+1}_{-1} - i163^{+10}_{-9}$  & $2540^{+6}_{-7} - i20^{+4}_{-4}$ & 2037 & 2488 & 2513 \\
F32P21/F48P21 & 208 & $2157^{+3}_{-3} - i120^{+7}_{-6}$ & $2497^{+6}_{-6} - i26^{+3}_{-3}$  & 2109 & 2457 & 2473  \\
F32P30/F48P30 & 305 & $ 2261^{+6}_{-11}$   & $2552^{+6}_{-7}-i34^{+4}_{-3}$  & 2271 & 2545 & 2562 \\
H48P32 & 317 & $2295^{+2}_{-3}$ & $2568^{+7}_{-8} - i33^{+4}_{-4}$  & 2296 & 2571 & 2588 \\
\bottomrule[2pt]
\end{tabular}
\label{tab:couple-pole}
\end{table}

We list in Table~\ref{tab:couple-pole} only the pole positions that lie close to the physical region. Compared with the results of the single-channel scheme in Table~\ref{tab:single-pole}, the poles on the second RS are well reproduced in the coupled-channel scheme. Moreover, for each ensemble, by taking the coupled-channel effects into account, an additional resonance state appears on the $[110]$ RS. This state lies significantly above the $D\pi$ threshold, and its real part is higher than the $D\eta$ threshold for $M_\pi\approx133$ and $208$ MeV (in the C48P14, F32P21, and F48P21 ensembles), but close to the $D\eta$ threshold for $M_\pi\approx305$ and 317 MeV (in the F32P30, F48P30, and H48P32 ensembles). {This state lies below the $D_s\bar K$ threshold in the F32P30, F48P30, and H48P32 ensembles, but above its threshold for the other ensembles.} 
Specifically, these resonances reside on an unphysical RS that is directly connected to the physical region and are expected to have a noticeable impact on physical observables. The results support the two-pole structure of the $D_0^*(2300)$ \cite{Kolomeitsev:2003ac,Guo:2006fu,Guo:2009ct,Guo:2015dha,Albaladejo:2016lbb}. Specifically, at the physical pion mass ($M_\pi\approx133$ MeV in the C48P14 ensemble), these two poles are located at $2149^{+1}_{-1} - i163^{+10}_{-9}$ and $2540^{+6}_{-7} - i20^{+4}_{-4}$ MeV, respectively, which is basically consistent with the findings of Ref.~\cite{Albaladejo:2016lbb}. In particular, with physical masses and $F_\pi$, the two poles are located at $2112^{+1}_{-1}-i155_{-8}^{+9}$ and $2491^{+7}_{-6}-i30^{+4}_{-4}$ MeV, respectively. 
The corresponding effective couplings $g_i$ of these two poles to the channel $i$, which can be used to study their decay properties, are summarized in Table~\ref{tab:pole:residue}. These couplings are obtained from the residue of the scattering amplitude $\mathbb{T}_{ij}$ in Eq.~\eqref{equ:Tcp} at the corresponding pole position: 
\bea
g_i g_j = \lim_{s\to s_\text{pole}}(s-s_\text{pole})\mathbb{T}_{ij}(s)\ .\label{equ:effective_coupling}
\eea
The calculated effective couplings in Table~\ref{tab:pole:residue} indicate that these two poles couple predominantly to the $D\pi$ and $D_s\bar{K}$ channels, respectively, which is consistent with previous results in Refs.~\cite{Albaladejo:2016lbb,Guo:2018tjx}.
\renewcommand{\arraystretch}{1}
\begin{table}[tb]
\caption{Central values of effective couplings of the two poles of the $D_0^*(2300)$ for different ensembles, as defined in Eq.~\eqref{equ:effective_coupling} for the $I=1/2$ $D\pi$-$D\eta$-$D_s\bar{K}$ coupled-channel system.}
\centering
\footnotesize
\begin{tabular}{c|c|ccccc}
\toprule[2pt]
\makebox[5em][c]{\text{ensemble}} & \makebox[3em][c]{$M_\pi$} & \makebox[4em][c]{RS}& \makebox[10em][c]{$E_{\rm pole}$} & \makebox[6em][c]{$\lvert g_1 \rvert~\text{(MeV)}$} & \makebox[5em][c]{$\lvert g_2/g_1 \rvert$} & \makebox[5em][c]{$\lvert g_3/g_1 \rvert$} \\
\midrule[0.5pt]
\multirow{2}{*}{C48P14}& \multirow{2}{*}{133} & $[100]$ &$2149 - i163$ & $10464$ & $0.09$ & $0.53$ \\
& & $[110]$ &$2540 - i20 $ & $3085$ & $2.50$ & $3.47$ \\[7.5pt]
\multirow{2}{*}{F32P21/F48P21} &\multirow{2}{*}{208} & $[100]$ &$2157 - i120$ & $10401$ & $0.05$ & $0.53$ \\
& & $[110]$ & $2497 - i26$& $3781$ & $2.05$ & $2.84$ \\[7.5pt]
\multirow{2}{*}{F32P30/F48P30} &\multirow{2}{*}{305} & $[100]$ &$2261$ & $10492$ & $0.004$ & $0.54$ \\
& & $[110]$ &$2552 -i 34$ & $4126$ & $1.84$ & $2.76$ \\[7.5pt]
\multirow{2}{*}{H48P32}&\multirow{2}{*}{317} & $[100]$ & $2295$& $6076$ & $0.01$ & $0.55$ \\
& & $[110]$ &$2568 - i33$ & $4189$ & $1.85$ & $2.85$ \\
\bottomrule[2pt]
\end{tabular}
\label{tab:pole:residue}
\end{table}

\begin{figure}[tb]
\begin{center}
\begin{tikzpicture}
    \node[anchor=south west,inner sep=0] (image) at (0,0) {
        \includegraphics[width=0.75\linewidth]{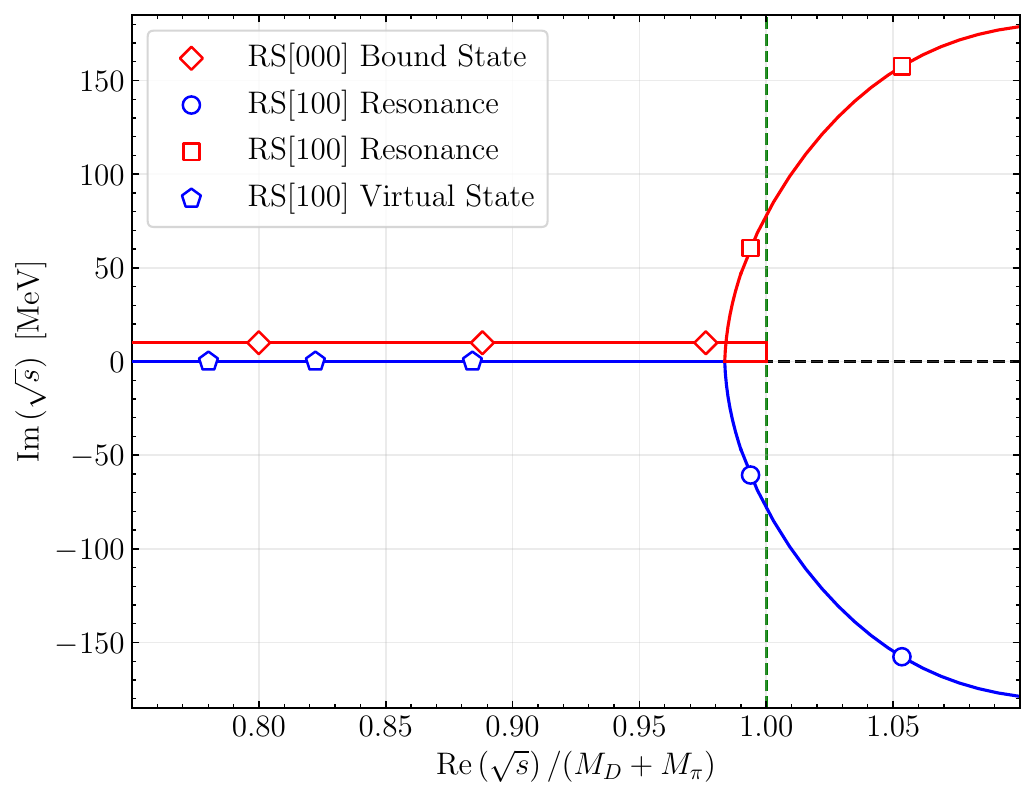}
    };
    
    \begin{scope}[x={(image.south east)},y={(image.north west)}]
       
       \draw[-{Stealth[open, length=8pt, width=6pt,fill=white]},thick,blue] (0.55,0.547) -- (0.54,0.547);
       \draw[-{Stealth[open, length=8pt, width=6pt,fill=white]},thick,blue] (0.365,0.547) -- (0.355,0.547);
       \draw[-{Stealth[open, length=8pt, width=6pt,fill=white]},thick,blue] (0.806,0.245) -- (0.800,0.255);

       \draw[-{Stealth[open, length=8pt, width=6pt,fill=white]},thick,red] (0.57,0.572) -- (0.56,0.572);
       \draw[-{Stealth[open, length=8pt, width=6pt,fill=white]},thick,red] (0.797,0.836) -- (0.789,0.823);
       \draw[-{Stealth[open, length=8pt, width=6pt,fill=white]},thick,red] (0.385,0.572) -- (0.375,0.572);
       \draw[-{Stealth[open, length=8pt, width=6pt,fill=white]},thick,red] (0.725,0.547) -- (0.735,0.547);

       \draw[-{Stealth[open, length=8pt, width=6pt,fill=white]},thick,black] (0.775,0.471) -- (0.707,0.531);
       \draw[-{Stealth[open, length=8pt, width=6pt,fill=white]},thick,black] (0.816,0.602) -- (0.748,0.558);
       \draw[-{Stealth[open, length=8pt, width=6pt,fill=white]},thick,black] (0.662,0.776) -- (0.734,0.726);

        \node[rotate=90, green!50!black, font=\scriptsize] at (0.72,0.24) {$D\pi$ threshold};
        \node[blue, font=\scriptsize] at (0.879,0.150) {1};
        \node[blue, font=\scriptsize] at (0.716,0.378) {2};
        \node[blue, font=\scriptsize] at (0.467,0.520) {3};
        \node[blue, font=\scriptsize] at (0.316,0.520) {4};
        \node[blue, font=\scriptsize] at (0.212,0.520) {5};
        \node[red, font=\scriptsize] at (0.852,0.937) {1};
        \node[red, font=\scriptsize] at (0.702,0.709) {2};
        \node[red, font=\scriptsize] at (0.666,0.596) {3};
        \node[red, font=\scriptsize] at (0.454,0.596) {4};
        \node[red, font=\scriptsize] at (0.236,0.596) {5};
        \node[black, font=\scriptsize] at (0.653,0.792) {$M_{\pi} = 262$ MeV};
        \node[black, font=\scriptsize] at (0.833,0.452) {$M_{\pi} = 299$ MeV};
        \node[black, font=\scriptsize] at (0.839,0.626) {$M_{\pi} = 327$ MeV};
    \end{scope}
\end{tikzpicture}
\caption{Pole trajectory of the $I = 1/2$ $D\pi$–$D\eta$–$D_s\bar{K}$ coupled-channel unitarized amplitude around 2200 MeV, where $n$ is defined by $M_\pi = n M_\pi^{\text{ph}}$.}
\label{fig:couple:trajectory}
\end{center}
\end{figure}
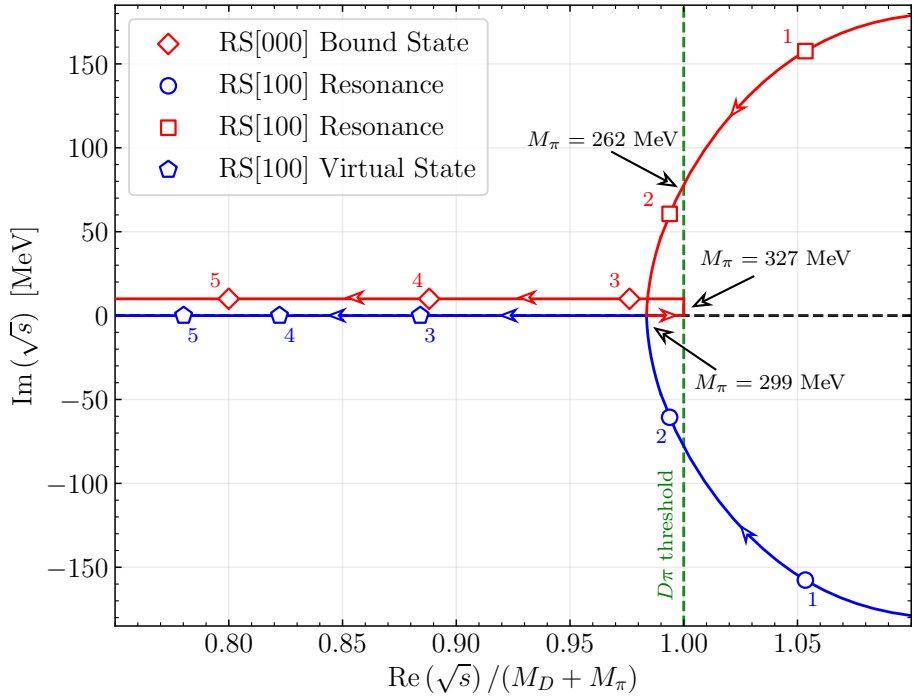

To investigate the trajectory of the pole position as a function of the pion mass within the coupled-channel scheme, we also require the pion mass dependence of the masses of the mesons involved in the coupled channels, namely $M_K$, $M_D$, $M_\eta$, and $M_{D_s}$. Analogous to Eq.~\eqref{equ:MD}, where we obtained the pion mass dependence of $M_D$, we employ the relation from Refs.~\cite{Gasser:1984gg,Guo:2015dha} for the kaon:
\begin{align}
    M_K^2 = \bar{M}_K^2+\frac12 M_\pi^2\ ,\label{eq:MK}
\end{align}
where the kaon mass in the chiral limit, $\bar{M}_K=484.6(6)\ \text{MeV}$, is obtained by fitting the values of $M_K$ from the various ensembles in Ref.~\cite{CLQCD:2023sdb} (as listed in Table~\ref{tab:setting}) to Eq.~\eqref{eq:MK}. The remaining pion mass dependence of $M_\eta$ and $M_{D_s}$ is then obtained using Eq.~\eqref{equ:Meta_MDs}. As in the single-channel fit, we select 25 pion mass points ranging from 30 to 700 MeV to compute the corresponding pole trajectory for the pole on the $[100]$ RS in Fig.~\ref{fig:couple:trajectory}. For $M_\pi$ between 30 MeV and 299 MeV, the poles correspond to the conjugate resonance states. As the pion mass increases, the ratio of the resonance mass to the $D\pi$ threshold gradually decreases, and the width becomes narrower. In the range $299~\text{MeV} < M_\pi < 327~\text{MeV}$, the poles transition into two virtual states. As the pion mass increases further, the ratio of the pole mass to the $D\pi$ threshold for one virtual state continues to decrease, while that for the other increases, reaching unity at $M_\pi = 327$ MeV. Beyond this point, the latter pole moves onto the physical RS and becomes a bound state. For $M_\pi > 327$ MeV, the ratios of the masses of the two poles (one virtual and one bound) to the $D\pi$ threshold decrease with the increase of pion mass. The behavior of the coupled-channel pole trajectory shown in Fig.~\ref{fig:couple:trajectory} is fully consistent with that of the single-channel case presented in Fig.~\ref{fig:single:trajectory}. Both are in good agreement with the description in Ref.~\cite{Guo:2015dha}, which is based on $D\pi$-$D\eta$-$D_s\bar{K}$ coupled-channel unitarized chiral amplitudes up to NLO. Furthermore, we employed a similar method to calculate the trajectory of the pole on the $[110]$ RS, as shown in Fig.~\ref{fig:couple:trajectory_110}.

\begin{figure}[tb]
\begin{center}
\begin{tikzpicture}
    \node[anchor=south west,inner sep=0] (image) at (0,0) {
        \includegraphics[width=0.75\linewidth]{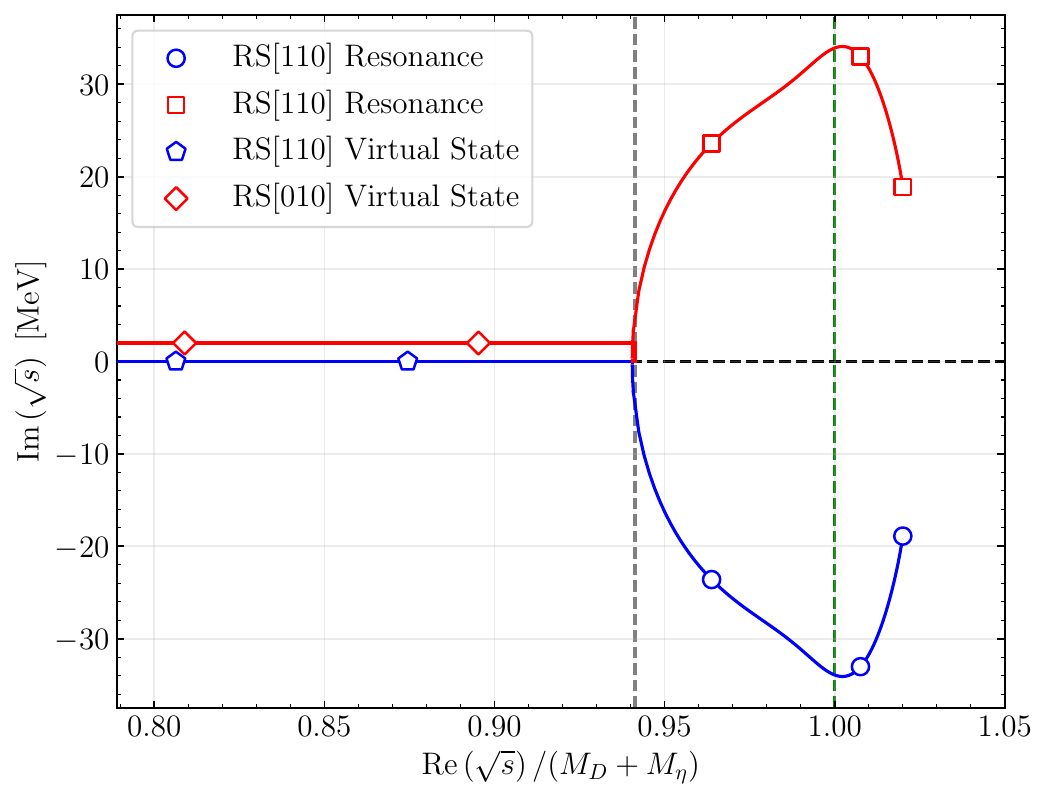}
    };
    
    \begin{scope}[x={(image.south east)},y={(image.north west)}]
 
       \draw[-{Stealth[open, length=8pt, width=6pt,fill=white]},thick,blue] (0.52,0.547) -- (0.51,0.547);
       \draw[-{Stealth[open, length=8pt, width=6pt,fill=white]},thick,blue] (0.335,0.547) -- (0.325,0.547);
       \draw[-{Stealth[open, length=8pt, width=6pt,fill=white]},thick,blue] (0.702,0.248) -- (0.692,0.258);

       \draw[-{Stealth[open, length=8pt, width=6pt,fill=white]},thick,red] (0.706,0.852) -- (0.695,0.839);
       \draw[-{Stealth[open, length=8pt, width=6pt,fill=white]},thick,red] (0.410,0.570) -- (0.4008,0.570);

       \draw[-{Stealth[open, length=8pt, width=6pt,fill=white]},thick,black] (0.675,0.471) -- (0.607,0.541);
       \draw[-{Stealth[open, length=8pt, width=6pt,fill=white]},thick,black] (0.525,0.471) -- (0.597,0.541);
       \draw[-{Stealth[open, length=8pt, width=6pt,fill=white]},thick,black] (0.71,0.908) -- (0.79,0.938);

        \node[rotate=90, green!50!black, font=\scriptsize] at (0.78,0.28) {$D\eta$ threshold};
        \node[blue, font=\scriptsize] at (0.879,0.300) {1};
        \node[blue, font=\scriptsize] at (0.846,0.150) {2};
        \node[blue, font=\scriptsize] at (0.650,0.258) {3};
        \node[blue, font=\scriptsize] at 
        (0.401,0.520) {4};
        \node[blue, font=\scriptsize] at (0.170,0.520) {5};
        \node[red, font=\scriptsize] at (0.879,0.797) {1};
        \node[red, font=\scriptsize] at (0.844,0.937) {2};
        \node[red, font=\scriptsize] at (0.676,0.846) {3};
        \node[red, font=\scriptsize] at (0.482,0.594) {4};
        \node[red, font=\scriptsize] at (0.196,0.594) {5};
        \node[black, font=\scriptsize] at (0.508,0.452) {$M_{\pi} = 460.8$ MeV};
        \node[black, font=\scriptsize] at (0.710,0.452) {$M_{\pi} = 461.3$ MeV};
        \node[black, font=\scriptsize] at (0.639,0.891) {$M_{\pi} = 317$ MeV};
    \end{scope}
\end{tikzpicture}
\caption{Pole trajectory of the $I = 1/2$ $D\pi$--$D\eta$--$D_s\bar{K}$ coupled-channel unitarized amplitude around $2500$~MeV, where $n$ is defined by $M_\pi = n M_\pi^{\text{ph}}$.  The vertical gray dashed line marks the $D\pi$ threshold at $M_\pi = 461.3$ MeV, where a virtual state on the $[110]$ sheet crosses into the connected $[010]$ sheet.}
\label{fig:couple:trajectory_110}
\end{center}
\end{figure}
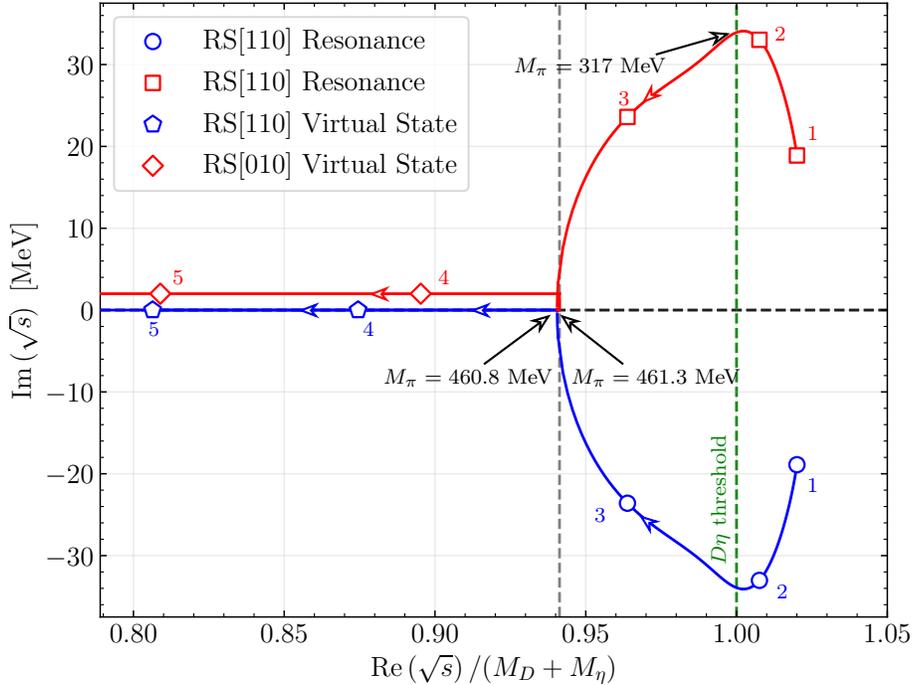

At last, we examine the impact of the finite lattice spacing $a$ (see Table~\ref{tab:setting}). Since the lattice spacing governs the ultraviolet behavior, its influence can be absorbed into the subtraction constant. Specifically, we extend Eq.~\eqref{eq:amu} by including an $a^2$ term:
\begin{equation}
a(\mu) = \bar{a}(\mu) + \beta M_\pi^2 + \gamma a^2,
\end{equation}
where $\gamma$ is a free parameter. However, our results indicate that the $\gamma a^2$ term has a negligible impact on the three fitting parameters listed in Table~\ref{tab:parameters}, and therefore does not alter our finite-volume results. Moreover, the inclusion of this additional parameter $\gamma$ increases the $\chi^2/\text{d.o.f.}$ values reported in Table~\ref{tab:paras:T} to 2.55 (single-channel) and 2.18 (coupled-channel), respectively.

\section{summary}\label{sec:summary}
We reanalyze the lattice spectra in the $A_1^+$ irreducible representation for $I=1/2$ $D\pi$ scattering presented in Ref.~\cite{Yan:2024yuq} to investigate the impact of chiral and SU(3) symmetries on $S$-wave $D\pi$ scattering and the $D_0^*(2300)$ resonance. First, we use the chirally modified ERE and $K$-matrix parameterizations, constructed by including an energy-dependent factor to preserve the correct chiral behavior following Ref.~\cite{Du:2025beb}, to fit the phase shifts extracted from the lattice spectra via L\"uscher's formula. We then compare the resulting pole positions with those obtained from the traditional parameterizations to highlight the effects of chiral symmetry. Furthermore, based on unitarized chiral perturbation theory (ChPT), we employ both the single-channel scheme and the $D\pi$–$D\eta$–$D_s\bar{K}$ coupled-channel scheme to directly describe the lattice spectra. In this analysis, we compare the fitting efficacy, pole positions, pole trajectory behavior, and effective couplings to demonstrate the combined effects of chiral symmetry and SU(3) flavor symmetry in studying $S$-wave $D\pi$ scattering and the $D_0^*(2300)$ resonance.

By comparing the traditional parameterization with the modified one, our results show that the chiral factor shifts the real part of the pole position closer to the threshold. Notably, for $M_\pi \approx 133$ MeV, the effect of chiral symmetry is more important: the chiral factor reduces the mass obtained from the traditional parameterization by approximately $200$ MeV and its width by approximately $400$ MeV. 

In the subsequent unitarized ChPT analysis, the types, positions, and numbers of these poles after including the chiral factor at different $M_\pi$ are further confirmed in both the single- and coupled-channel schemes. In particular, by considering the chiral extrapolation in ChPT, we study the behavior of the pole trajectory in the unitarized chiral amplitude as a function of $M_\pi$. We also find that, even when using the same number of free parameters in the fit, the $D\pi$–$D\eta$–$D_s\bar{K}$ coupled-channel scheme yields noticeably better results than the single-channel approach. {This improvement is attributed to the fact that the lattice simulations are performed on symmetric $N_f = 2 + 1$ Wilson-Clover ensembles, where sea-quark effects naturally necessitate the inclusion of the $D\eta$ and $D_s\bar{K}$ channels, even though the two-hadron operators considered in Ref.~\cite{Yan:2024yuq} are limited to $D^{(*)}\pi$}.
Notably, once the coupled-channel effect is incorporated, a resonance state with a mass of approximately $2500$ MeV is found on the $[110]$ RS. This resonance pole is expected to have a noticeable impact on physical observables and supports the two-pole structure of the $D_0^*(2300)$. Specifically, near the physical pion mass ($M_\pi \approx 133$ MeV), these two poles are located at $2149^{+1}_{-1} - i163^{+10}_{-9}$ and $2540^{+6}_{-7} - i20^{+4}_{-4}$ MeV, respectively. In particular, with physical masses and $F_\pi$, the corresponding two poles are located at $2112^{+1}_{-1}-i155_{-8}^{+9}$ and $2491^{+7}_{-6}-i30^{+4}_{-4}$ MeV, respectively. The calculated effective couplings indicate that these two poles couple predominantly to the $D\pi$ and $D_s\bar{K}$ channels, respectively, which is consistent with previous results.

These results indicate that chiral symmetry can significantly affects the extracted mass and width of a state. {Within this context, incorporating an energy-dependent chiral factor to the ERE and $K$-matrix parameterizations can largely reproduce the effects of unitarized ChPT and improve the reliability of the extracted pole positions, especially for resonances. This further highlights the role of chiral symmetry as a guiding principle in constructing reliable low-energy parameterizations of the system with the pseudoscalar mesons treated as Goldstone bosons.} Moreover, the results demonstrate that the coupled-channel effect, required by SU(3) symmetry, is crucial for generating the two-pole structure of the $D_0^*(2300)$.

\begin{acknowledgements}
We are grateful to Feng-Kun Guo, Liuming Liu and Zhengli Wang for useful discussions. Finite-volume energy levels taken from Ref.~\cite{Yan:2024yuq} were provided by the China Lattice QCD Collaboration (CLQCD). No endorsement on their part of the analysis presented in the current paper should be
assumed. This work is supported in part by the National Natural Science Foundation of China (NSFC) under Grant No.~12547111; by the Generalitat Valenciana (GVA) under the PROMETEU program with Ref. CIPROM/2023/59; and by MICIU/AEI/10.13039/501100011033 under grants PID2023-147458NB-C21 and CEX2023-001292-S.
\end{acknowledgements}

\bibliography{ref.bib}
 
\end{document}